\documentclass[reprint,superscriptaddress,aps,nofootinbib,preprintnumbers]{revtex4-2}
\usepackage{amsmath}
\usepackage{amssymb}
\usepackage{amsthm}
\usepackage{mathtools}
\usepackage{physics}
\usepackage{bm}
\usepackage{bbm}
\usepackage[caption=false]{subfig}
\usepackage{booktabs}
\usepackage{graphicx}
\usepackage{tikz}
\usepackage{comment}
\usepackage{bbold}
\usepackage[colorlinks, allcolors=blue]{hyperref}

\newcommand{\site}{\bm{r}}
\newcommand{\link}[1][ij]{\langle #1 \rangle}
\newcommand{\Cm}[1][]{C_{\text{m}#1}}
\newcommand{\Cg}[1][]{C_{\text{g}#1}}
\newcommand{\CJ}[1][]{C_{\text{J}#1}}
\newcommand{\EJ}[1][]{E_{\text{J}#1}}
\newcommand{\EC}[1][]{E_{\text{C}#1}}
\newcommand{\cen}[1][n]{\mathrm{ce}_{#1}}
\newcommand{\sen}[1][n]{\mathrm{se}_{#1}}

\newcommand{\Proj}{\mathbb{P}}
\newcommand{\Sproj}{\mathbb{S}}
\newcommand{\Id}{\mathbbm{1}}

\newcommand{\EHUQC}{EHU Quantum Center and Department of Physical Chemistry, University of the Basque Country UPV/EHU, P.O. Box 644, 48080 Bilbao, Spain}
\newcommand{\BCMaterials}{BCMaterials, Basque Center for Materials, Applications and Nanostructures, UPV/EHU Science Park, 48940 Leioa, Spain}
\newcommand{\DIPC}{DIPC - Donostia International Physics Center, Paseo Manuel de Lardizabal 4, 20018 San Sebastián, Spain}
\newcommand{\IKERBASQUE}{IKERBASQUE, Basque Foundation for Science, Plaza Euskadi 5, 48009 Bilbao, Spain}
\newcommand{\CERN}{European Organization for Nuclear Research (CERN),  Theoretical Physics Department, CH-1211 Geneva, Switzerland}
\newcommand{\IQC}{Institute for Quantum Computing and Department of Electrical \& Computer Engineering, University of Waterloo, Waterloo, Ontario, N2L 3G1, Canada}

\begin{document}

\title{Compact U(1) Lattice Gauge Theory in Superconducting Circuits with Infinite-Dimensional Local Hilbert Spaces}

\author{J. M. Alcaine-Cuervo}
\email{jesusmatias.alcaine@ehu.eus}
\affiliation{\EHUQC}
\affiliation{\BCMaterials}

\author{S. Pradhan}
\email{sunny.pradhan@ehu.eus}
\affiliation{\EHUQC}
\affiliation{\DIPC}

\author{E. Rico}
\email{enrique.rico.ortega@gmail.com}
\affiliation{\EHUQC}
\affiliation{\DIPC}
\affiliation{\IKERBASQUE}
\affiliation{\CERN}

\author{Z. Shi}
\email{zheng.shi@uwaterloo.ca}
\affiliation{\IQC}

\author{C.M. Wilson}
\email{chris.wilson@uwaterloo.ca}
\affiliation{\IQC}

\preprint{CERN-TH-2026-003}

\begin{abstract}
We propose a superconducting-circuit architecture that realizes a compact $\mathrm{U(1)}$ lattice gauge theory using the intrinsic infinite-dimensional Hilbert space of phase and charge variables.
The gauge and matter fields are encoded directly in the degrees of freedom of the rotor variables associated with the circuit nodes, and Gauss’s law emerges exactly from the conservation of local charge, without auxiliary stabilizers, penalty terms, or Hilbert-space truncation.
A minimal gauge–matter coupling arises microscopically from Josephson nonlinearities, whereas the magnetic plaquette interaction is generated perturbatively via virtual matter excitations.
Numerical diagonalization confirms the emergence of compact electrodynamics and coherent vortex excitations, underscoring the need for large local Hilbert spaces in the continuum regime.
The required circuit parameters are within the current experimental capabilities.
Our results establish superconducting circuits as a scalable, continuous-variable platform for analog quantum simulation of non-perturbative gauge dynamics.
\end{abstract}

\maketitle

\section{Introduction}
\label{sec:sec_1}

Quantum simulators for lattice gauge theories (LGTs)~\cite{Banuls2020b,Alexeev2021,Bauer2023b,DiMeglio2024} provide a promising route to explore nonperturbative phenomena such as confinement, real-time string breaking, and topological excitations~\cite{De2024,Liu2025,Luo2025,Mildenberger2025,Zhang2025,Alexandrou2025,Schuhmacher2025,Xiang2025,Halimeh2025a}, recently expanding to (2+1)D systems ~\cite{Zohar2022,Meth2025,Crippa2024,Gyawali2025,Cochran2025,GonzalezCuadra2025,Cobos2025}.
Classical approaches, including tensor networks \cite{Banuls2019,Banuls2020a,Magnifico2025} and Monte Carlo methods~\cite{Kogut1979,Creutz1979,Rebbi1983}, face severe limitations in real-time or finite-density regimes, motivating the development of hardware platforms capable of directly implementing gauge degrees of freedom (d.o.f).

Superconducting circuits offer high coherence, controllable interactions, and native access to conjugate phase-charge variables~\cite{Clarke2008,Devoret2013,Wendin2017,Krantz2019,Kjaergaard2020,Blais2021, Kwon2021, Gao2021, Rasmussen2021}, making them an attractive platform for gauge-theory simulation~\cite{Marcos2013,Marcos2014,Brennen2016,Sameti2017,Alaeian2019,Klco2020,Atas2021,Homeier2021,Hung2021,Pedersen2021,Ciavarella2021,Satzinger2021,Ge2022,Wang2022,Kane2022,Pardo2023,Belyansky2024,Busnaina2024,Charles2024,Wang2025b}.
Existing proposals, however, predominantly employ qubits~\cite{Mazzola2021,Busnaina2025} or finite-level truncations~\cite{Popov2024,Jakobs2025,Mirandariaza2025}, obscuring essential features of compact $\mathrm{U(1)}$ theories, including the continuous rotor Hilbert space, the full operator algebra, and the structure of vortex excitations~\cite{Zhang2023,Pardo2025,Lin2025}.
Approaches based on continuous-variable encoding avoid explicit truncation but require manual enforcement of compactness~\cite{Stavenger2022,Ale2025}.

In this work, we introduce a superconducting-circuit realization of compact $\mathrm{U(1)}$ LGT in which both compactness and gauge invariance arise naturally from the microscopic circuit structure.
Gauge and matter fields are encoded via rotor variables associated with circuit nodes and links, exploiting the intrinsic infinite-dimensional Hilbert space of superconducting phase and charge.
A crucial aspect of this implementation is that Gauss's law follows from Kirchhoff's current conservation and circuit topology, rather than being an imposed constraint.
Additionally, the minimal gauge-matter coupling arises directly from the nonlinearity of the Josephson junction, without auxiliary elements.

We show that, in the regime where matter excitations are virtual, an effective magnetic plaquette interaction is generated at fourth order in perturbation theory, reproducing the Kogut--Susskind Hamiltonian of compact quantum electrodynamics (QED).
Exact diagonalization of a single plaquette confirms the emergence of vortex excitations and demonstrates the necessity of large local Hilbert spaces to access the continuum regime.
The required circuit parameters lie well within the reach of state-of-the-art superconducting devices, enabling a scalable route to analog quantum simulation of non-perturbative gauge dynamics in two dimensions.

\section{The compact U(1) model}
\label{sec:sec_2}

We consider the Kogut--Susskind Hamiltonian formulation of compact $\mathrm{U(1)}$ gauge theory on a square lattice \cite{PhysRevD.11.395}.
Gauge fields reside on links $(\site, \hat{\mu})$, where $\site$ indicates a lattice site and $\hat{\mu} = \hat{x},\hat{y}$ a spatial direction.
These fields are represented by rotor variables $\hat{U}_{\site, \mu} = \exponential(i \hat{\theta}_{\site, \mu} )$, where $\hat{\theta}_{\site, \mu} \in [-\pi, \pi)$ is the gauge phase, and their conjugate variables are the electric fields $\hat{E}_{\site, \mu}$, which have integer eigenvalues.
They satisfy the commutation relationship $[ \hat{E}_{\site, \mu}, \hat{U}_{\site', \nu}] = - \delta_{\site, \site^{\prime}} \delta_{\mu, \nu} \hat{U}_{\site, \mu}$.

The pure-gauge Hamiltonian can be written as
\begin{equation}
    \hat{H}_{\text{g}}  = g^2 \sum_{(\site, \mu)} \hat{E}_{\site, \mu}^2 + \frac{1}{2 g^2} \sum_{\site} \left[ \hat{P}_{\site} + \hat{P}_{\site}^\dagger \right],
    \label{eq:pure_gauge_hamiltonian}
\end{equation}
where $\hat{P}(\site) = \hat{U}_{\site, x} \hat{U}_{\site + \hat{x}, y} \hat{U}^{\dagger}_{\site + \hat{y}, x} \hat{U}^{\dagger}_{\site, y}$.
We define the \emph{plaquette operator} $\hat{\square}_{\site} \equiv (\hat{P}_{\site} + \hat{P}_{\site}^{\dagger}) / 2$, which in terms of the gauge phase $\hat{\theta}_{\site}$ reads
\begin{equation}
    \hat{\square}_{\site}
    = \cos(\hat{\theta}_{\site, x} + \hat{\theta}_{\site + \hat{x}, y} - \hat{\theta}_{\site + \hat{y}, x} - \hat{\theta}_{\site, y}),
    \label{eq:plaq_operator_LGT}
\end{equation}
highlighting the model's compact nature.

Bosonic matter fields $\exponential(i \hat{\phi}_{\site})$ reside on lattice sites and satisfy $[\hat{n}_{\site}, \exponential(i \hat{\phi}_{\site'})] = - \delta_{\site, \site^{\prime}} \exponential(i \hat{\phi}_{\site})$, where $\hat{n}_{\site}$ is the occupation number operator at site $\site$.
The matter Hamiltonian consists of a mass term and a minimal gauge-matter coupling,~
\begin{equation}
    \hat{H}_\text{m} = m  \sum_{\site} \hat{n}_{\site}^2 - \lambda \sum_{\site, \mu} \cos(\hat{\phi}_{\site} + \hat{\theta}_{\site, \mu} - \hat{\phi}_{\site + \hat{\mu}}).
\end{equation}

The full Hamiltonian of the compact QED model is just the sum $H_{\rm QED} = H_{\text{g}} + H_{\text{m}}$.
Local gauge transformations are generated by the operators
\begin{equation}
    \hat{G}_{\site} = \hat{n}_{\site} - \sum_{\mu} \left( \hat{E}_{\site, \mu} - \hat{E}_{\site - \hat{\mu}, \mu} \right) ,
    \label{eq:gauge_transf_op}
\end{equation}
which commute with the Hamiltonian, i.e.~$[\hat{G}_{\site}, \hat{H}] = 0$.
Furthermore, physical states must satisfy $\hat{G}_{\site} \ket{\Psi} = q_{\site} \ket{\Psi}$ for each $\site$, where $q_{\site}$ denotes a distribution of static charges.
These constraints are equivalent to a discretized form of Gauss's law.

\section{Superconducting circuits realization}
\label{sec:sc_circuit}

A superconducting circuit is characterized by a set of phase operators $\hat{\phi}_j$ and their conjugate charge operators $\hat{n}_{\phi_j}$, satisfying $[\exponential(i \hat{\phi}_j),\, \hat{n}_k] = \delta_{jk} \exponential(i \hat{\phi}_j)$.
In the phase basis, $\hat{n}_{\phi_j}=- i \partial_{\phi_j}$ (in units of $2e$).
On the other hand, the exponential operators $\exponential(\pm i\hat{\phi})$ shift the integer-valued charge number and realize the $\mathrm{U(1)}$ rotor algebra.
These variables therefore provide a natural framework for constructing a fully $\mathrm{U(1)}$ gauge-invariant circuit model.

We start by considering a simple circuit composed of two nodes with phases $\hat{\phi}_i$ and $\hat{\phi}_j$, connected to a common ground and linked through a third node with phase $\hat{\theta}_{ij}$.
The first two nodes encode matter d.o.f while the intermediate one encodes a gauge field d.o.f.
A Josephson junction on the link generates the interaction $\hat{H}_{{\rm J},ij }= - E_{\rm J} \cos{(\hat{\phi}_i + \hat{\theta}_{ij} - \hat{\phi}_{j})}$,
which implements the minimal gauge–matter coupling.
This setup represents only one side of the full plaquette, and its Hamiltonian reads
\begin{equation}
    \hat{H}_{ij} =
    m \, (\hat{n}_{\phi_i}^2 + \hat{n}_{\phi_j}^2  )
    + g \, \hat{n}_{\theta_{ij}}^2
    - \lambda \cos{(\hat{\phi}_i + \hat{\theta}_{ij} - \hat{\phi}_j)} ,
    \label{eq:minimal_hamiltonian}
\end{equation}
where $m \equiv 2e^2 / \Cm $ and $g \equiv 2 e^2 / \Cg$ are the charging energies associated with the matter and gauge capacitances $\Cm$ and $\Cg$, respectively, and $\lambda \equiv \EJ$ is the Josephson energy.
Parasitic junction capacitances $\CJ$ generate additional cross-terms that renormalize $m$ and $g$, while preserving gauge invariance due to the underlying circuit symmetry.
For a more detailed discussion, see Appendix~\ref{app_sec:Circuit_quantization}. Throughout this work, we consider a homogeneous circuit with $\Cg[,ij]=\Cg,\ \Cm[,i]=\Cm,\ \EJ[,ij]=\EJ$.

\begin{figure}[t]
    \centering
    \includegraphics[scale=0.26]{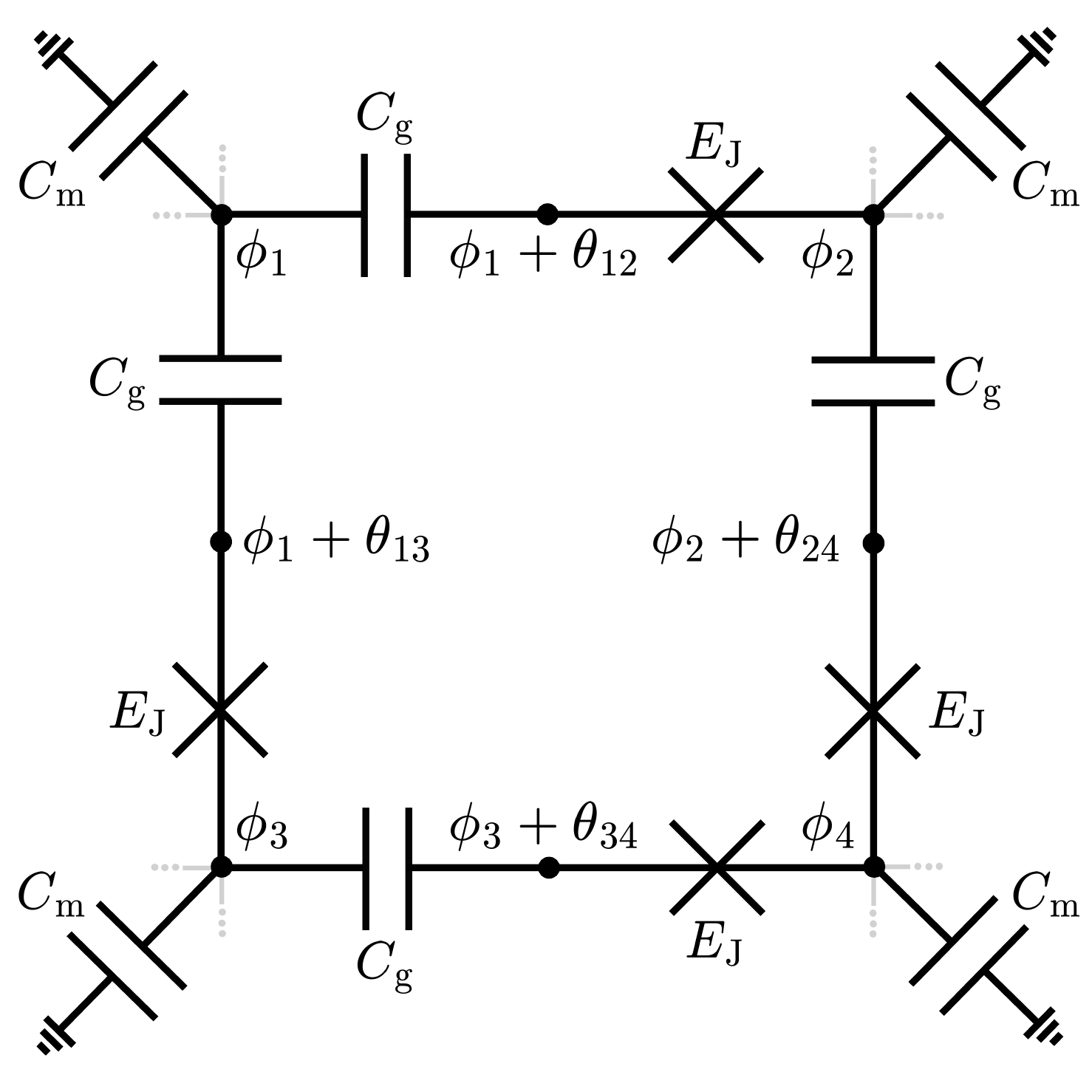}
    \caption{Lumped-element superconducting-circuit realization of a compact $\mathrm{U(1)}$ plaquette.
    Matter d.o.f are encoded in site-node phases $\hat{\phi}_i$ and conjugate charges $\hat{n}_i$, while gauge fields reside on link-node phases $\hat{\theta}_{ij}$ and charges $\hat{n}_{ij}$. Josephson junctions generate the minimal gauge–matter coupling $\cos{( \hat{\phi}_i+\hat{\theta}_{ij} - \hat{\phi}_j )}$. Gauss’s law emerges exactly from charge conservation at each node and requires no auxiliary constraints.}
    \label{fig:Plaquette}
\end{figure}

Connecting four such gauge–matter circuit branches into a closed loop [see Fig.~\ref{fig:Plaquette}] yields the plaquette Hamiltonian~
\begin{equation}
    \hat{H}_{\mathrm{P}} = m \sum_i  \hat{n}^2_i + g \sum_{\link} \hat{n}^2_{ij} - \lambda \sum_{\link}  \cos{( \hat{\phi}_i + \hat{\theta}_{ij} - \hat{\phi}_j )},
    \label{eq:plaquette_hamiltonian}
\end{equation}
where $\langle ij \rangle \in \{12,24,13,34\}$ labels the links, $\hat{n}_i \equiv \hat{n}_{\phi_i}$ and $\hat{n}_{ij} \equiv \hat{n}_{\theta_{ij}}$.
Charge conservation at each matter node $i$ is given by the Gauss operators
\begin{equation}
    \hat{G}_i = \hat{n}_i - \sum_{\link} s_{ij} \hat{n}_{ij},
    \label{eq:circuit_gauss_laws}
\end{equation}
which are the circuit counterparts of the generators of gauge transformations \eqref{eq:gauge_transf_op} in compact $\mathrm{U(1)}$ LGT. Here, the sum runs over all links connected to site $i$, and $s_{ij}=\pm1$ denotes their orientation. These operators correspond to conserved quantities of the circuit (see Appendix~\ref{app:Gauss_match_dof}). Since both the capacitive and Josephson terms satisfy this constraint, they commute with the Hamiltonian, i.e.~$[\hat{G}_i, \hat{H}] = 0$, ensuring that Gauss’s law is preserved at every site.

We restrict our analysis to the sector with vanishing static charges, $q_i=0$, corresponding to the vacuum of the gauge model.
Physical states must therefore satisfy the condition $\hat{G}_i \ket{\psi} = 0$ for all vertices $i$ and, as a result, these states have redundant d.o.f.
The extra d.o.f can be eliminated by gauge-fixing, for instance by expressing the matter charges $\hat n_i$ in terms of the gauge ones $\hat{n}_{ij}$.
While this parametrization is not unique, we adopt it as we consider it the most symmetric one.

For the numerical study of the system, the charge basis becomes inefficient in the regime $\lambda/m, \lambda/g \gg 1$, where the cosine terms dominate. In the limit $m=0$, the plaquette \eqref{eq:plaquette_hamiltonian} decomposes into four independent link Hamiltonians, whose eigenfunctions are given by Mathieu functions~\cite{Cottet2002, Didier2018, Koch2007, Olver2010, Arfken2013} (see Appendix~\ref{app_sec:Mathieu}). We employ this Mathieu basis to diagonalize the Hamiltonian for $m\neq0$.

\begin{figure}[t]
    \centering
    \includegraphics[width = \columnwidth]{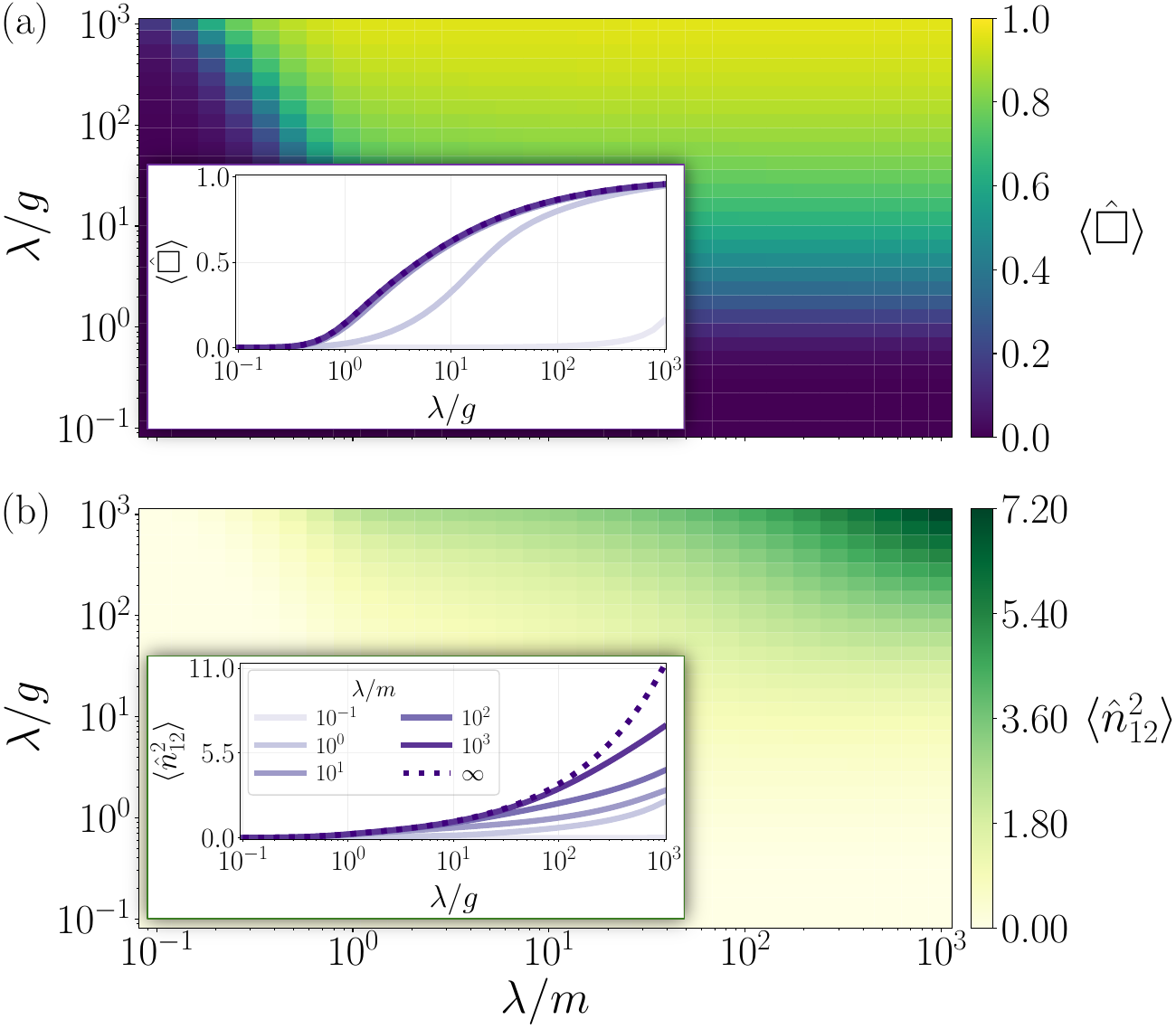}
    \caption{Ground-state properties of the plaquette Hamiltonian. (a) Expectation value of the plaquette operator $ \langle \hat{\square}\rangle$ as a function of the couplings $\lambda / g$ and $\lambda / m$. (b) Electric-field fluctuations $\langle \hat{n}_{12}^2 \rangle$ on a representative link. Large values of $\lambda / g$, corresponding to the weak-coupling (continuum) limit of compact electrodynamics, require large local Hilbert spaces, highlighting the necessity of untruncated rotor d.o.f. Insets show cuts at fixed values of $\lambda /m$ and share a common legend. Data are obtained for the local Hilbert space of dimension $N = 13$. Numerical convergence is discussed in Appendix~\ref{app_sec:Numerical_results}.
}
    \label{fig:GS_observables}
\end{figure}

The circuit equivalent of the plaquette operator \eqref{eq:plaq_operator_LGT} is given by~
\begin{equation}
    \hat{\square} = \cos{(\hat{\theta}_{12} + \hat{\theta}_{24} - \hat{\theta}_{13} - \hat{\theta}_{34})}.
    \label{eq:plaq_operator_circuit}
\end{equation}
In Fig.~\ref{fig:GS_observables} we show its expectation value on the ground state, together with the corresponding electric-field fluctuations, as functions of $\lambda/g$ and $\lambda/m$.
For $\lambda/g \ll1$, electric-field fluctuations are suppressed and the system is dominated by the electric-field energy. In contrast, for $\lambda/g \gg 1$, corresponding to the continuum limit of compact QED, the plaquette expectation value approaches unity, and vortex excitations become energetically relevant. Achieving accurate convergence in this regime requires large local Hilbert spaces, underscoring a fundamental limitation of truncated gauge-field encodings. Comparable behavior is observed in tensor network simulations of a four-plaquette lattice (see Appendix~\ref{app_sec:Numerical_results}).

\section{Effective Plaquette Hamiltonian}
\label{sec:sec_4}

In a LGT, the plaquette operator $\hat{\square}$ is fundamental because it introduces dynamics for the gauge fields. This term, however, is explicitly missing in Eq.~\eqref{eq:plaquette_hamiltonian}.
Nonetheless, it can be recovered when $m \gg \lambda, g$, which we label as the \emph{static matter regime}.
Here, matter excitations are suppressed in the low-energy sector, and the ground state is characterized by the absence of charges on the matter nodes.
Furthermore, the cosine potential creates a matter–antimatter charge pair on a plaquette edge, with a gauge charge between them.
Consequently, in the static matter regime, the cosine potentials can excite the gauge links around the plaquette, while leaving the matter nodes untouched, if it goes through three ``virtual'' states with matter excitations.

We can deduce that the energy scale of such processes is $O(\lambda^4 / m^3)$.
Therefore, a perturbative analysis up to the fourth order can reveal if the plaquette term does, in fact, emerge.
As shown in full detail in Appendix~\ref{app_sec:Heff_static_m}, the effective Hamiltonian up to fourth order in $\lambda$ reads
\begin{equation}
    \hat{H}_{\text{eff}} = g \sum_{\langle ij \rangle}\hat{n}^2_{ij} + J_\square \cos{(\hat{\theta}_{12} + \hat{\theta}_{24} - \hat{\theta}_{13} - \hat{\theta}_{34})}
    \label{eq:effective_Hamiltonian}
\end{equation}
where $J_\square =- 5 \lambda^4 / (16m^3)$ is the coupling of the plaquette term.
This effective Hamiltonian is restricted to the subspace of states with $\hat{n}_i = 0$ for all matter nodes $i$.
We have also estimated that this perturbative expansion is well-defined as long as $ \lambda < m / 8$, which gives a domain of validity of Eq.~\eqref{eq:effective_Hamiltonian}.
Exact diagonalization confirms excellent agreement between the full circuit Hamiltonian and this effective low-energy gauge theory within the expected regime (see Appendix~\ref{app_sec:Numerical_results}).

\section{Vortex States: Preparation and Detection}
\label{sec:sec_5}

In compact $\mathrm{U(1)}$ gauge theories, vortex excitations encode topological sectors associated with large gauge transformations rather than merely local defects.
On the lattice, a vortex threading a plaquette implements a quantized magnetic flux, representing a discrete analog of the continuum $\Theta$-vacuum structure.
The coherent preparation and manipulation of such vortex states therefore provide direct experimental access to topological interference and vacuum-structure effects in compact gauge theories.

Vortex states correspond to eigenstates of the plaquette operator.
Such states can be found in the regime $\lambda/g\gg 1$ with $\lambda/m\gtrsim  1$, where the vacuum expectation of the plaquette operator satisfies $\langle \hat{\square}\rangle \sim 1$~\cite{Bauer2023, Haase2021}.
A vortex state can be generated with the operator~
\begin{equation}
    \hat{U}_{\text{vortex}}(\Theta)
     = \exp(\frac{i \Theta}{4} (\hat{n}_{12}+\hat{n}_{24} - \hat{n}_{34} - \hat{n}_{13})),
     \label{eq:vortex_op}
\end{equation}
which displaces the flux combination $\hat{\theta}_{12}+\hat{\theta}_{24} - \hat{\theta}_{34} - \hat{\theta}_{13}$ by $-\Theta$.
This linear combination of phases corresponds to the total flux going around a plaquette and, in the limit $m \to \infty$, it can be identified with the only effective d.o.f of the plaquette (see Appendix \ref{app_sec:Mathieu} for more details).
Acting with $\hat{U}_{\text{vortex}}$ on the vacuum $\ket{\Psi_0}$ of the system produces the state $\ket{\Theta} \equiv \hat{U}_{\text{vortex}} (\Theta) \ket{\Psi_0}$.
Experimentally, $\ket{\Theta}$ can be prepared via a sudden change of the gate voltage applied simultaneously on all links of the plaquette loop.
In Sec.~\ref{sec:sec_6} we further discuss the readout of vortex dynamics.

\begin{figure}[t]
    \centering
    \includegraphics[width = \columnwidth]{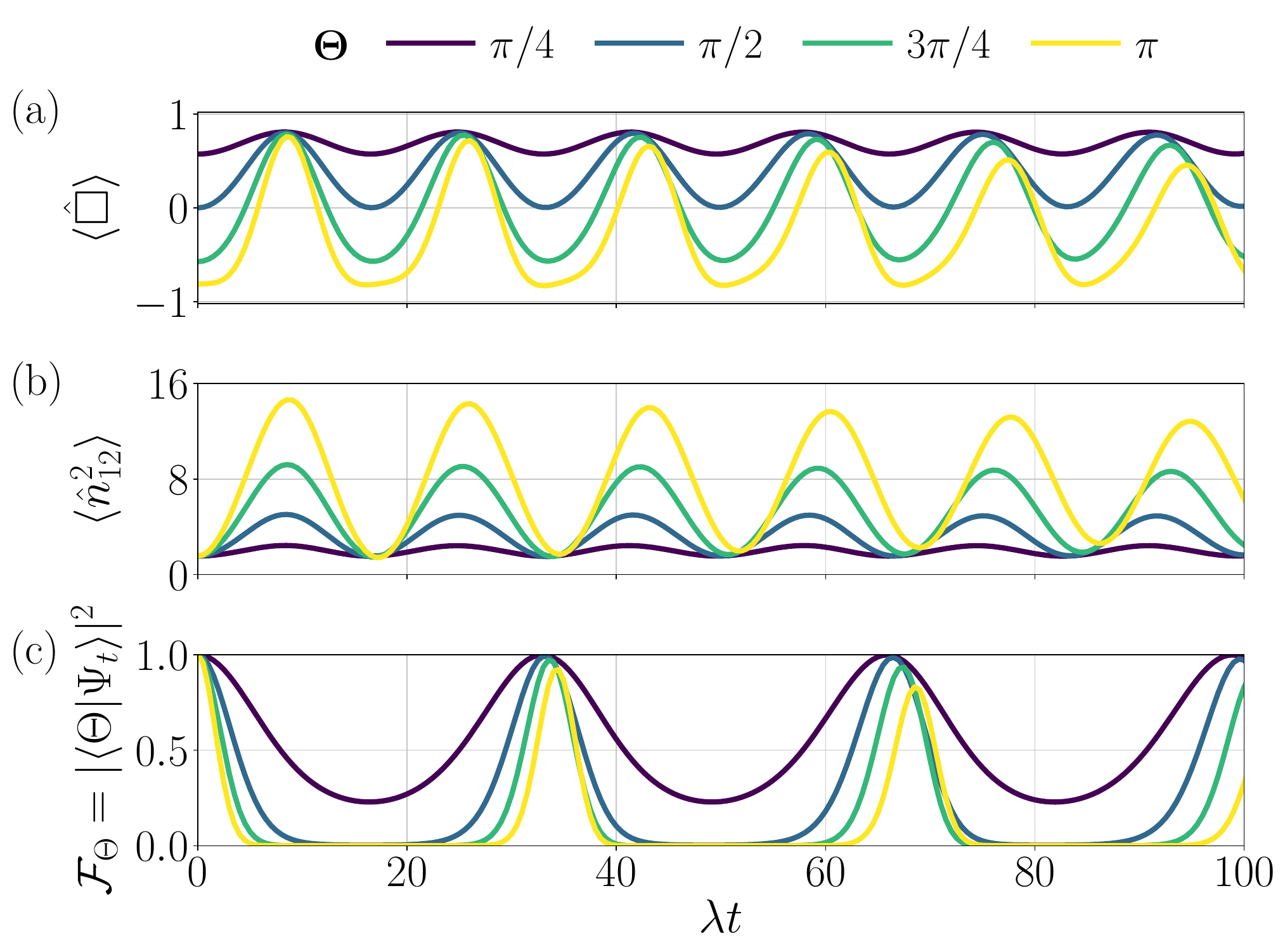}
    \caption{Coherent vortex dynamics on a single plaquette.
    Time evolution of (a) the plaquette operator $\langle \hat{\square}\rangle $ in \eqref{eq:plaq_operator_circuit}, (b) the electric-field variance $\langle \hat{n}^2_{12} \rangle$, and (c) the fidelity $\mathcal{F}_{\Theta}  = |\langle \Theta | \Psi_{t} \rangle|^2$.
    As an initial state, we have considered a vortex state prepared by applying $\hat{U}_{\text{vortex}}$ in \eqref{eq:vortex_op} to the ground state of the plaquette Hamiltonian \eqref{eq:plaquette_hamiltonian}.
    Persistent oscillations indicate coherent vortex dynamics accessible with experimentally realistic parameters.
    Results are shown for $\lambda / m = \lambda / g = 50$ and local Hilbert space of dimension $N = 13$.
}
\label{fig:vortex_tevol}
\end{figure}

Fig. \ref{fig:vortex_tevol} shows coherent real-time dynamics of vortex states generated by flux-threading operators. The persistence of oscillations in both magnetic and electric observables indicates coherent vortex dynamics on experimentally relevant timescales. Details of the numerical implementation are provided in Appendix~\ref{app_sec:Numerical_results} .

\section{Experimental Feasibility}
\label{sec:sec_6}

Designing a circuit is a constrained optimization between materials, geometry, and aspects like the desired coherence times \cite{LevensonFalk2025}.
A wide range of capacitive energy scales, $\EC/h \sim 0.003 - 30$~GHz, are used in superconducting circuits~\cite{Nakamura1999, Martinis2002, Krantz2019, Blais2021}.
Together with dc SQUIDs where the effective $\EJ$ can be tuned from practically zero to tens of gigahertz~\cite{Nakamura1999} (see Appendix~\ref{app_sec:Circuit_quantization}), this covers both rotor-like and transmon-like regimes in our work, where $\lambda \sim \EJ$ and $m, g \sim \EC$.
Effective couplings $|J_\square|/h$ can be as large as $10$~MHz, well within decoherence rate $\Gamma_1/(2\pi)$, $\Gamma_2/(2\pi) \le 10$~kHz of state-of-the-art devices.

The four-junction plaquette is compatible with standard fabrication and layout techniques, and the symmetry of the capacitances ensures robust gauge invariance.
Scaling to multiplaquette arrays requires only extending the circuit graph. No stabilizers or energy penalties are needed, making the extension to larger lattices straightforward.

The vortex dynamics shown in Fig.~\ref{fig:vortex_tevol} have timescales of the order of nanoseconds.
Their experimental detection may be simplest in the frequency domain, where the gigahertz Josephson radiation emitted by the circuit can be measured~\cite{Pedersen1976, Soerensen1977, karimi2024}.
It should also be possible to directly detect the time dynamics using nonadiabatic control techniques, such as those used in early experiments with charge qubits~\cite{Nakamura1999}.
Even 25 years ago, those experiments used $100$~ps control pulses to measure dynamics on the scale of $\EJ/h\sim20$~GHz.
In this method, after the initial state is prepared and coherent vortex dynamics are turned on for a variable time, $\lambda$ is nonadiabatically tuned to small values.
This arrests the vortex dynamics such that the charge states on all links can be read out.

\section{Conclusions and outlook}
\label{sec:sec_7}

In this work, we have introduced a superconducting-circuit architecture that realizes compact $\mathrm{U(1)}$ LGT with continuous rotor d.o.f.
An immediate advantage of our setup is the implementation of gauge invariance, a crucial aspect of LGTs simulation.
It arises naturally thanks to Kirchhoff's laws and the circuit topology.
Moreover, the encoding of the LGT d.o.f with the circuit variables avoids any kind of truncation, which allows us to explore phenomenological aspects that require a large local Hilbert space.
Furthermore, we have demonstrated that an effective magnetic interaction can arise from controlled fourth-order virtual processes.
This enables the observation of vortex physics and nonperturbative gauge dynamics and, in fact, we have shown that it is possible to prepare states that display coherent vortex dynamics.  
Our scheme operates in experimentally accessible regimes, and the generalization to multiplaquette setups is straightforward.
Therefore, we can conclude that the presented circuit setup provides a scalable platform for analog quantum simulation of $\mathrm{U(1)}$ gauge theories.

\begin{acknowledgments}
    We thank I.L. Egusquiza for discussions and comments on the manuscript. J.M.A.-C. and E.R. acknowledge the financial support received from the IKUR Strategy under the collaboration agreement between the Ikerbasque Foundation and UPV/EHU on behalf of the Department of Education of the Basque Government.
    E.R. acknowledges support from the BasQ strategy of the Department of Science, Universities, and Innovation of the Basque Government.
    E.R. is supported by the grant PID2021-126273NB-I00 funded by MCIN/AEI/10.13039/501100011033 and by “ERDF A way of making Europe” and the Basque Government through Grant No. IT1470-22.

    This work has been partially funded by the Eric \& Wendy Schmidt Fund for Strategic Innovation through the CERN Next Generation Triggers project under grant agreement number SIF-2023-004.

    CMW and ZS acknowledge the Canada First Research Excellence Fund (CFREF), NSERC of Canada, the Canadian Foundation for Innovation, the Ontario Ministry of Research and Innovation, and Industry Canada for financial support.
\end{acknowledgments}

\bibliography{bibliography}

@article{Banuls2020b,
  title={Simulating lattice gauge theories within quantum technologies},
  author={Bañuls, Mari Carmen and Blatt, Rainer and Catani, Jacopo and Celi, Alessio and Cirac, Juan Ignacio and Dalmonte, Marcello and Fallani, Leonardo and Jansen, Karl and Lewenstein, Maciej and Montangero, Simone and Muschik, Christine A. and Reznik, Benni and Rico, Enrique and Tagliacozzo, Luca and Van Acoleyen, Karel and Verstraete, Frank and Wiese, Uwe-Jens and Wingate, Matthew and Zakrzewski, Jakub and Zoller, Peter},
  journal={Eur. Phys. J. D},
  volume={74},
  number={8},
  pages={165},
  year={2020},
  publisher={Springer},
  doi={10.1140/epjd/e2020-100571-8}
}

@article{Alexeev2021,
  title = {Quantum Computer Systems for Scientific Discovery},
  author = {Alexeev, Yuri and Bacon, Dave and Brown, Kenneth R. and Calderbank, Robert and Carr, Lincoln D. and Chong, Frederic T. and DeMarco, Brian and others},
  journal = {PRX Quantum},
  volume = {2},
  issue = {1},
  pages = {017001},
  numpages = {19},
  year = {2021},
  month = {Feb},
  publisher = {American Physical Society},
  doi = {10.1103/PRXQuantum.2.017001},
  url = {https://link.aps.org/doi/10.1103/PRXQuantum.2.017001}
}

@article{Bauer2023b,
  title = {Quantum Simulation for High-Energy Physics},
  author = {Bauer, Christian W. and Davoudi, Zohreh and Balantekin, A. Baha and Bhattacharya, Tanmoy and Carena, Marcela and de Jong, Wibe A. and Draper, Patrick and others},
  journal = {PRX Quantum},
  volume = {4},
  issue = {2},
  pages = {027001},
  numpages = {70},
  year = {2023},
  month = {May},
  publisher = {American Physical Society},
  doi = {10.1103/PRXQuantum.4.027001},
  url = {https://link.aps.org/doi/10.1103/PRXQuantum.4.027001}
}

@article{DiMeglio2024,
  title = {Quantum Computing for High-Energy Physics: State of the Art and Challenges},
  author = {Di Meglio, Alberto and Jansen, Karl and Tavernelli, Ivano and Alexandrou, Constantia and Arunachalam, Srinivasan and Bauer, Christian W. and Borras, Kerstin and Carrazza, Stefano and Crippa, Arianna and Croft, Vincent and de Putter, Roland and Delgado, Andrea and Dunjko, Vedran and Egger, Daniel J. and Fern\'andez-Combarro, Elias and Fuchs, Elina and Funcke, Lena and Gonz\'alez-Cuadra, Daniel and Grossi, Michele and Halimeh, Jad C. and Holmes, Zo\"e and K\"uhn, Stefan and Lacroix, Denis and Lewis, Randy and Lucchesi, Donatella and Martinez, Miriam Lucio and Meloni, Federico and Mezzacapo, Antonio and Montangero, Simone and Nagano, Lento and Pascuzzi, Vincent R. and Radescu, Voica and Ortega, Enrique Rico and Roggero, Alessandro and Schuhmacher, Julian and Seixas, Joao and Silvi, Pietro and Spentzouris, Panagiotis and Tacchino, Francesco and Temme, Kristan and Terashi, Koji and Tura, Jordi and T\"uys\"uz, Cenk and Vallecorsa, Sofia and Wiese, Uwe-Jens and Yoo, Shinjae and Zhang, Jinglei},
  journal = {PRX Quantum},
  volume = {5},
  issue = {3},
  pages = {037001},
  numpages = {49},
  year = {2024},
  month = {Aug},
  publisher = {American Physical Society},
  doi = {10.1103/PRXQuantum.5.037001},
  url = {https://link.aps.org/doi/10.1103/PRXQuantum.5.037001}
}

@misc{De2024,
      title={Observation of string-breaking dynamics in a quantum simulator}, 
      author={Arinjoy De and Alessio Lerose and De Luo and Federica M. Surace and Alexander Schuckert and Elizabeth R. Bennewitz and Brayden Ware and William Morong and Kate S. Collins and Zohreh Davoudi and Alexey V. Gorshkov and Or Katz and Christopher Monroe},
      year={2024},
      eprint={2410.13815},
      archivePrefix={arXiv},
      primaryClass={quant-ph},
      url={https://arxiv.org/abs/2410.13815}, 
}

@article{Liu2025,
  title = {String-Breaking Mechanism in a Lattice Schwinger Model Simulator},
  author = {Liu, Ying and Zhang, Wei-Yong and Zhu, Zi-Hang and He, Ming-Gen and Yuan, Zhen-Sheng and Pan, Jian-Wei},
  journal = {Phys. Rev. Lett.},
  volume = {135},
  issue = {10},
  pages = {101902},
  numpages = {8},
  year = {2025},
  month = {Sep},
  publisher = {American Physical Society},
  doi = {10.1103/mwy1-v9hk},
  url = {https://link.aps.org/doi/10.1103/mwy1-v9hk}
}

@misc{Luo2025,
      title={Quantum simulation of bubble nucleation across a quantum phase transition}, 
      author={De Luo and Federica Maria Surace and Arinjoy De and Alessio Lerose and Elizabeth R. Bennewitz and Brayden Ware and Alexander Schuckert and Zohreh Davoudi and Alexey V. Gorshkov and Or Katz and Christopher Monroe},
      year={2025},
      eprint={2505.09607},
      archivePrefix={arXiv},
      primaryClass={quant-ph},
      url={https://arxiv.org/abs/2505.09607}, 
}

@article{Mildenberger2025,
   title={Confinement in a ${{\mathbb{Z}}}_{2}$ lattice gauge theory on a quantum computer},
   volume={21},
   ISSN={1745-2481},
   url={http://dx.doi.org/10.1038/s41567-024-02723-6},
   DOI={10.1038/s41567-024-02723-6},
   number={2},
   journal={Nat. Phys.},
   publisher={Springer Science and Business Media LLC},
   author={Mildenberger, Julius and Mruczkiewicz, Wojciech and Halimeh, Jad C. and Jiang, Zhang and Hauke, Philipp},
   year={2025},
   month=jan, pages={312–317} 
}

@article{Zhang2025,
  title={Observation of microscopic confinement dynamics by a tunable topological $\theta$-angle},
  author={Wei-Yong Zhang and Ying Liu and Yanting Cheng and Ming-Gen He and Han-Yi Wang and Tian-Yi Wang and Zi-Hang Zhu and Guo-Xian Su and Zhao-Yu Zhou and Yong-Guang Zheng and Hui Sun and Bing Yang and Philipp Hauke and Wei Zheng and Jad C. Halimeh and Zhen-Sheng Yuan and Jian-Wei Pan},
  journal={Nat. Phys.},
  volume={21},
  number={1},
  pages={155--160},
  year={2025},
  publisher={Nature Publishing Group UK London},
  doi={10.1038/s41567-024-02702-x}
}

@misc{Alexandrou2025,
      title={Realizing string breaking dynamics in a ${{\mathbb{Z}
}}_{2}$ lattice gauge theory on quantum hardware}, 
      author={Constantia Alexandrou and Andreas Athenodorou and Kostas Blekos and Georgios Polykratis and Stefan Kühn},
      year={2025},
      eprint={2504.13760},
      archivePrefix={arXiv},
      primaryClass={hep-lat},
      url={https://arxiv.org/abs/2504.13760}, 
}

@misc{Schuhmacher2025,
      title={Observation of hadron scattering in a lattice gauge theory on a quantum computer}, 
      author={Julian Schuhmacher and Guo-Xian Su and Jesse J. Osborne and Anthony Gandon and Jad C. Halimeh and Ivano Tavernelli},
      year={2025},
      eprint={2505.20387},
      archivePrefix={arXiv},
      primaryClass={quant-ph},
      url={https://arxiv.org/abs/2505.20387}, 
}

@misc{Xiang2025,
      title={Real-time scattering and freeze-out dynamics in Rydberg-atom lattice gauge theory}, 
      author={De-Sheng Xiang and Peng Zhou and Chang Liu and Hao-Xiang Liu and Yao-Wen Zhang and Dong Yuan and Kuan Zhang and Biao Xu and Marcello Dalmonte and Dong-Ling Deng and Lin Li},
      year={2025},
      eprint={2508.06639},
      archivePrefix={arXiv},
      primaryClass={cond-mat.quant-gas},
      url={https://arxiv.org/abs/2508.06639}, 
}

@misc{Halimeh2025a,
      title={Quantum simulation of out-of-equilibrium dynamics in gauge theories}, 
      author={Jad C. Halimeh and Niklas Mueller and Johannes Knolle and Zlatko Papić and Zohreh Davoudi},
      year={2025},
      eprint={2509.03586},
      archivePrefix={arXiv},
      primaryClass={quant-ph},
      url={https://arxiv.org/abs/2509.03586}, 
}

@article{Zohar2022,
    author = {Zohar, Erez},
    title = {Quantum simulation of lattice gauge theories in more than one space dimension—requirements, challenges and methods},
    journal = {Philos. Trans. R. Soc. A},
    volume = {380},
    number = {2216},
    pages = {20210069},
    year = {2021},
    month = {12},
    issn = {1364-503X},
    doi = {10.1098/rsta.2021.0069},
    url = {https://doi.org/10.1098/rsta.2021.0069}
}

@article{Meth2025,
  author       = {Michael Meth and Jinglei Zhang and Jan F. Haase and Claire Edmunds and Lukas Postler and Alexander Steiner and Luca Dellantonio and Rainer Blatt and Peter Zoller and Thomas Monz and Philipp Schindler and Christine Muschik and Martin Ringbauer},
  title        = {Simulating two-dimensional lattice gauge theories on a qudit quantum computer},
  journal      = {Nat. Phys.},
  year         = {2025},
  volume       = {21},
  issue        = {4},
  pages        = {570--576},
  doi          = {10.1038/s41567-025-02797-w},
}

@misc{Crippa2024,
      title={Analysis of the confinement string in (2 + 1)-dimensional Quantum Electrodynamics with a trapped-ion quantum computer}, 
      author={Arianna Crippa and Karl Jansen and Enrico Rinaldi},
      year={2024},
      eprint={2411.05628},
      archivePrefix={arXiv},
      primaryClass={hep-lat},
      url={https://arxiv.org/abs/2411.05628}, 
}

@misc{Gyawali2025,
      title={Observation of disorder-free localization using a (2+1)D lattice gauge theory on a quantum processor}, 
      author={Gaurav Gyawali and Shashwat Kumar and Yuri D. Lensky and Eliott Rosenberg and Aaron Szasz and Tyler Cochran and Renyi Chen and Amir H. Karamlou and Kostyantyn Kechedzhi and Julia Berndtsson and Tom Westerhout and others},
      year={2025},
      eprint={2410.06557},
      archivePrefix={arXiv},
      primaryClass={quant-ph},
      url={https://arxiv.org/abs/2410.06557}, 
}

@article{Cochran2025,
   title={Visualizing dynamics of charges and strings in (2 + 1)D lattice gauge theories},
   volume={642},
   ISSN={1476-4687},
   url={http://dx.doi.org/10.1038/s41586-025-08999-9},
   DOI={10.1038/s41586-025-08999-9},
   number={8067},
   journal={Nature},
   publisher={Springer Science and Business Media LLC},
   author={Cochran, T. A. and Jobst, B. and Rosenberg, E. and Lensky, Y. D. and Gyawali, G. and Eassa, N. and Will, M. and Szasz, A. and others},
   year={2025},
   month=jun, pages={315–320} 
}

@article{GonzalezCuadra2025,
   title={Observation of string breaking on a (2 + 1)D Rydberg quantum simulator},
   volume={642},
   ISSN={1476-4687},
   url={http://dx.doi.org/10.1038/s41586-025-09051-6},
   DOI={10.1038/s41586-025-09051-6},
   number={8067},
   journal={Nature},
   publisher={Springer Science and Business Media LLC},
   author={González-Cuadra, Daniel and Hamdan, Majd and Zache, Torsten V. and Braverman, Boris and Kornjača, Milan and Lukin, Alexander and Cantú, Sergio H. and Liu, Fangli and Wang, Sheng-Tao and Keesling, Alexander and Lukin, Mikhail D. and Zoller, Peter and Bylinskii, Alexei},
   year={2025},
   month=jun, pages={321–326} 
}

@misc{Cobos2025,
      title={Real-Time Dynamics in a (2+1)-D Gauge Theory: The Stringy Nature on a Superconducting Quantum Simulator}, 
      author={Jesús Cobos and Joana Fraxanet and César Benito and Francesco di Marcantonio and Pedro Rivero and Kornél Kapás and Miklós Antal Werner and Örs Legeza and Alejandro Bermudez and Enrique Rico},
      year={2025},
      eprint={2507.08088},
      archivePrefix={arXiv},
      primaryClass={quant-ph},
      url={https://arxiv.org/abs/2507.08088}, 
}

@article{Banuls2019,
  author = "Bañuls, Mari Carmen  and  Cichy, Krzysztof  and  Cirac, J. Ignacio  and  Jansen, Karl  and  Kühn, Stefan",
  title = "{Tensor Networks and their use for Lattice Gauge Theories}",
  doi = "10.22323/1.334.0022",
  journal = "PoS",
  year = 2019,
  volume = "LATTICE2018",
  pages = "022"
}

@article{Banuls2020a,
   title={Review on novel methods for lattice gauge theories},
   volume={83},
   ISSN={1361-6633},
   url={http://dx.doi.org/10.1088/1361-6633/ab6311},
   DOI={10.1088/1361-6633/ab6311},
   number={2},
   journal={Rep. Prog. Phys.},
   publisher={IOP Publishing},
   author={Carmen Bañuls, Mari and Cichy, Krzysztof},
   year={2020},
   month=jan, pages={024401} 
}

@article{Magnifico2025,
  author       = {Giuseppe Magnifico and Giovanni Cataldi and Marco Rigobello and Peter Majcen and Daniel Jaschke and Pietro Silvi and Simone Montangero},
  title        = {Tensor networks for lattice gauge theories beyond one dimension},
  journal      = {Commun. Phys.},
  year         = {2025},
  volume       = {8},
  number       = {1},
  pages        = {322},
  doi          = {10.1038/s42005-025-02125-x},
}

@article{Kogut1979,
  title = {An introduction to lattice gauge theory and spin systems},
  author = {Kogut, John B.},
  journal = {Rev. Mod. Phys.},
  volume = {51},
  issue = {4},
  pages = {659--713},
  numpages = {0},
  year = {1979},
  month = {Oct},
  publisher = {American Physical Society},
  doi = {10.1103/RevModPhys.51.659},
  url = {https://link.aps.org/doi/10.1103/RevModPhys.51.659}
}

@article{Creutz1979,
  title = {Monte Carlo study of Abelian lattice gauge theories},
  author = {Creutz, Michael and Jacobs, Laurence and Rebbi, Claudio},
  journal = {Phys. Rev. D},
  volume = {20},
  issue = {8},
  pages = {1915--1922},
  numpages = {0},
  year = {1979},
  month = {Oct},
  publisher = {American Physical Society},
  doi = {10.1103/PhysRevD.20.1915},
  url = {https://link.aps.org/doi/10.1103/PhysRevD.20.1915}
}

@book{Rebbi1983,
  title={Lattice gauge theories and Monte Carlo simulations},
  author={Rebbi, Claudio},
  year={1983},
  publisher={World scientific}
}

@article{Clarke2008,
  author = {John Clarke and Frank K. Wilhelm},
  title = {Superconducting Quantum Bits},
  journal = {Nature},
  volume = {453},
  number = {7198},
  pages = {1031--1042},
  year = {2008},
  doi = {10.1038/nature07128}
}

@article{Devoret2013,
  author = {Michel H. Devoret and Robert J. Schoelkopf},
  title = {Superconducting Circuits for Quantum Information: An Outlook},
  journal = {Science},
  volume = {339},
  number = {6124},
  pages = {1169--1174},
  year = {2013},
  doi = {10.1126/science.1231930}
}

@article{Krantz2019,
  author = {Peter Krantz and Morten Kjaergaard and Fei Yan and Terry P. Orlando and Simon Gustavsson and William D. Oliver},
  title = {A Quantum Engineer’s Guide to Superconducting Qubits},
  journal = {Applied Physics Reviews},
  volume = {6},
  number = {2},
  pages = {021318},
  year = {2019},
  doi = {10.1063/1.5089550}
}

@article{Kjaergaard2020,
  author = {Morten Kjaergaard and Michael E. Schwartz and Jan Braum{\"u}ller and Peter Krantz and J. I.-J. Wang and Simon Gustavsson and William D. Oliver},
  title = {Superconducting Qubits: Current State of Play},
  journal = {Annual Review of Condensed Matter Physics},
  volume = {11},
  number = {1},
  pages = {369--395},
  year = {2020},
  doi = {10.1146/annurev-conmatphys-031119-050605}
}

@article{Blais2021,
  author = {Alexandre Blais and Arne L. Grimsmo and Steven M. Girvin and Andreas Wallraff},
  title = {Circuit Quantum Electrodynamics},
  journal = {Reviews of Modern Physics},
  volume = {93},
  number = {2},
  pages = {025005},
  year = {2021},
  doi = {10.1103/RevModPhys.93.025005}
}

@article{Marcos2013,
  title = {Superconducting Circuits for Quantum Simulation of Dynamical Gauge Fields},
  author = {Marcos, D. and Rabl, P. and Rico, E. and Zoller, P.},
  journal = {Phys. Rev. Lett.},
  volume = {111},
  issue = {11},
  pages = {110504},
  numpages = {6},
  year = {2013},
  month = {Sep},
  publisher = {American Physical Society},
  doi = {10.1103/PhysRevLett.111.110504},
  url = {https://link.aps.org/doi/10.1103/PhysRevLett.111.110504}
}

@article{Marcos2014,
title = {Two-dimensional lattice gauge theories with superconducting quantum circuits},
journal = {Annals of Physics},
volume = {351},
pages = {634-654},
year = {2014},
issn = {0003-4916},
doi = {https://doi.org/10.1016/j.aop.2014.09.011},
url = {https://www.sciencedirect.com/science/article/pii/S0003491614002711},
author = {D. Marcos and P. Widmer and E. Rico and M. Hafezi and P. Rabl and U.-J. Wiese and P. Zoller}
}

@article{Brennen2016,
  title = {Loops and Strings in a Superconducting Lattice Gauge Simulator},
  author = {Brennen, G. K. and Pupillo, G. and Rico, E. and Stace, T. M. and Vodola, D.},
  journal = {Phys. Rev. Lett.},
  volume = {117},
  issue = {24},
  pages = {240504},
  numpages = {7},
  year = {2016},
  month = {Dec},
  publisher = {American Physical Society},
  doi = {10.1103/PhysRevLett.117.240504},
  url = {https://link.aps.org/doi/10.1103/PhysRevLett.117.240504}
}

@article{Sameti2017,
  title = {Superconducting quantum simulator for topological order and the toric code},
  author = {Sameti, Mahdi and Poto\ifmmode \check{c}\else \v{c}\fi{}nik, Anton and Browne, Dan E. and Wallraff, Andreas and Hartmann, Michael J.},
  journal = {Phys. Rev. A},
  volume = {95},
  issue = {4},
  pages = {042330},
  numpages = {20},
  year = {2017},
  month = {Apr},
  publisher = {American Physical Society},
  doi = {10.1103/PhysRevA.95.042330},
  url = {https://link.aps.org/doi/10.1103/PhysRevA.95.042330}
}

@article{Alaeian2019,
  title = {Creating lattice gauge potentials in circuit QED: The bosonic Creutz ladder},
  author = {Alaeian, Hadiseh and Chang, Chung Wai Sandbo and Moghaddam, Mehran Vahdani and Wilson, Christopher M. and Solano, Enrique and Rico, Enrique},
  journal = {Phys. Rev. A},
  volume = {99},
  issue = {5},
  pages = {053834},
  numpages = {10},
  year = {2019},
  month = {May},
  publisher = {American Physical Society},
  doi = {10.1103/PhysRevA.99.053834},
  url = {https://link.aps.org/doi/10.1103/PhysRevA.99.053834}
}

@article{Klco2020,
  title = {SU(2) non-Abelian gauge field theory in one dimension on digital quantum computers},
  author = {Klco, Natalie and Savage, Martin J. and Stryker, Jesse R.},
  journal = {Phys. Rev. D},
  volume = {101},
  issue = {7},
  pages = {074512},
  numpages = {10},
  year = {2020},
  month = {Apr},
  publisher = {American Physical Society},
  doi = {10.1103/PhysRevD.101.074512},
  url = {https://link.aps.org/doi/10.1103/PhysRevD.101.074512}
}

@article{Atas2021,
   title={SU(2) hadrons on a quantum computer via a variational approach},
   volume={12},
   ISSN={2041-1723},
   url={http://dx.doi.org/10.1038/s41467-021-26825-4},
   DOI={10.1038/s41467-021-26825-4},
   number={1},
   journal={Nat. Commun.},
   publisher={Springer Science and Business Media LLC},
   author={Atas, Yasar Y. and Zhang, Jinglei and Lewis, Randy and Jahanpour, Amin and Haase, Jan F. and Muschik, Christine A.},
   year={2021},
   month=nov 
}

@article{Homeier2021,
  title = {${{\mathbb{Z}}}_{2}$ lattice gauge theories and Kitaev's toric code: A scheme for analog quantum simulation},
  author = {Homeier, Lukas and Schweizer, Christian and Aidelsburger, Monika and Fedorov, Arkady and Grusdt, Fabian},
  journal = {Phys. Rev. B},
  volume = {104},
  issue = {8},
  pages = {085138},
  numpages = {19},
  year = {2021},
  month = {Aug},
  publisher = {American Physical Society},
  doi = {10.1103/PhysRevB.104.085138},
  url = {https://link.aps.org/doi/10.1103/PhysRevB.104.085138}
}

@article{Pedersen2021,
  title = {Lattice gauge theory and dynamical quantum phase transitions using noisy intermediate-scale quantum devices},
  author = {Pedersen, Simon Panyella and Zinner, Nikolaj Thomas},
  journal = {Phys. Rev. B},
  volume = {103},
  issue = {23},
  pages = {235103},
  numpages = {17},
  year = {2021},
  month = {Jun},
  publisher = {American Physical Society},
  doi = {10.1103/PhysRevB.103.235103},
  url = {https://link.aps.org/doi/10.1103/PhysRevB.103.235103}
}

@article{Ciavarella2021,
  title = {Trailhead for quantum simulation of SU(3) Yang-Mills lattice gauge theory in the local multiplet basis},
  author = {Ciavarella, Anthony and Klco, Natalie and Savage, Martin J.},
  journal = {Phys. Rev. D},
  volume = {103},
  issue = {9},
  pages = {094501},
  numpages = {45},
  year = {2021},
  month = {May},
  publisher = {American Physical Society},
  doi = {10.1103/PhysRevD.103.094501},
  url = {https://link.aps.org/doi/10.1103/PhysRevD.103.094501}
}

@article{Satzinger2021,
   title={Realizing topologically ordered states on a quantum processor},
   volume={374},
   ISSN={1095-9203},
   url={http://dx.doi.org/10.1126/science.abi8378},
   DOI={10.1126/science.abi8378},
   number={6572},
   journal={Science},
   publisher={American Association for the Advancement of Science (AAAS)},
   author={Satzinger, K. J. and Liu, Y.-J and Smith, A. and Knapp, C. and Newman, M. and Jones, C. and Chen, Z. and Quintana, C. and Mi, X. and Dunsworth, A. and Gidney, C. and others},
   year={2021},
   month=dec, pages={1237–1241} 
}

@article{Hung2021,
  title = {Quantum Simulation of the Bosonic Creutz Ladder with a Parametric Cavity},
  author = {Hung, Jimmy S. C. and Busnaina, J. H. and Chang, C. W. Sandbo and Vadiraj, A. M. and Nsanzineza, I. and Solano, E. and Alaeian, H. and Rico, E. and Wilson, C. M.},
  journal = {Phys. Rev. Lett.},
  volume = {127},
  issue = {10},
  pages = {100503},
  numpages = {7},
  year = {2021},
  month = {Sep},
  publisher = {American Physical Society},
  doi = {10.1103/PhysRevLett.127.100503},
  url = {https://link.aps.org/doi/10.1103/PhysRevLett.127.100503}
}

@article{Ge2022,
doi = {10.1088/1674-1056/ac380e},
url = {https://doi.org/10.1088/1674-1056/ac380e},
year = {2022},
month = {feb},
publisher = {Chinese Physical Society and IOP Publishing Ltd},
volume = {31},
number = {2},
pages = {020304},
author = {Ge, Zi-Yong and Huang, Rui-Zhen and Meng, Zi-Yang and Fan, Heng},
title = {Quantum simulation of lattice gauge theories on superconducting circuits: Quantum phase transition and quench dynamics},
journal = {Chin. Phys. B}
}

@article{Wang2022,
  title = {Observation of emergent ${{\mathbb{Z}}}_{2}$ gauge invariance in a superconducting circuit},
  author = {Wang, Zhan and Ge, Zi-Yong and Xiang, Zhongcheng and Song, Xiaohui and Huang, Rui-Zhen and Song, Pengtao and Guo, Xue-Yi and Su, Luhong and Xu, Kai and Zheng, Dongning and Fan, Heng},
  journal = {Phys. Rev. Res.},
  volume = {4},
  issue = {2},
  pages = {L022060},
  numpages = {6},
  year = {2022},
  month = {Jun},
  publisher = {American Physical Society},
  doi = {10.1103/PhysRevResearch.4.L022060},
  url = {https://link.aps.org/doi/10.1103/PhysRevResearch.4.L022060}
}

@misc{Kane2022,
      title={Efficient quantum implementation of 2+1 U(1) lattice gauge theories with Gauss law constraints}, 
      author={Christopher Kane and Dorota M. Grabowska and Benjamin Nachman and Christian W. Bauer},
      year={2022},
      eprint={2211.10497},
      archivePrefix={arXiv},
      primaryClass={quant-ph},
      url={https://arxiv.org/abs/2211.10497}, 
}

@article{Pardo2023,
  title = {Resource-efficient quantum simulation of lattice gauge theories in arbitrary dimensions: Solving for Gauss's law and fermion elimination},
  author = {Pardo, Guy and Greenberg, Tomer and Fortinsky, Aryeh and Katz, Nadav and Zohar, Erez},
  journal = {Phys. Rev. Res.},
  volume = {5},
  issue = {2},
  pages = {023077},
  numpages = {16},
  year = {2023},
  month = {May},
  publisher = {American Physical Society},
  doi = {10.1103/PhysRevResearch.5.023077},
  url = {https://link.aps.org/doi/10.1103/PhysRevResearch.5.023077}
}

@article{Belyansky2024,
  title = {High-Energy Collision of Quarks and Mesons in the Schwinger Model: From Tensor Networks to Circuit QED},
  author = {Belyansky, Ron and Whitsitt, Seth and Mueller, Niklas and Fahimniya, Ali and Bennewitz, Elizabeth R. and Davoudi, Zohreh and Gorshkov, Alexey V.},
  journal = {Phys. Rev. Lett.},
  volume = {132},
  issue = {9},
  pages = {091903},
  numpages = {10},
  year = {2024},
  month = {Feb},
  publisher = {American Physical Society},
  doi = {10.1103/PhysRevLett.132.091903},
  url = {https://link.aps.org/doi/10.1103/PhysRevLett.132.091903}
}

@article{Busnaina2024,
  title={Quantum simulation of the bosonic Kitaev chain},
  author={Busnaina, Jamal H and Shi, Zheng and McDonald, Alexander and Dubyna, Dmytro and Nsanzineza, Ibrahim and Hung, Jimmy SC and Chang, CW Sandbo and Clerk, Aashish A and Wilson, Christopher M},
  journal={Nat. Commun.},
  volume={15},
  number={1},
  pages={3065},
  year={2024},
  publisher={Nature Publishing Group UK London},
  doi={10.1038/s41467-024-47186-8}
}

@article{Charles2024,
  title = {Simulating ${{\mathbb{Z}}}_{2}$ lattice gauge theory on a quantum computer},
  author = {Charles, Clement and Gustafson, Erik J. and Hardt, Elizabeth and Herren, Florian and Hogan, Norman and Lamm, Henry and Starecheski, Sara and Van de Water, Ruth S. and Wagman, Michael L.},
  journal = {Phys. Rev. E},
  volume = {109},
  issue = {1},
  pages = {015307},
  numpages = {19},
  year = {2024},
  month = {Jan},
  publisher = {American Physical Society},
  doi = {10.1103/PhysRevE.109.015307},
  url = {https://link.aps.org/doi/10.1103/PhysRevE.109.015307}
}

@misc{Wang2025b,
      title={Observation of Inelastic Meson Scattering in a Floquet System using a Digital Quantum Simulator}, 
      author={Ziting Wang and Zi-Yong Ge and Yun-Hao Shi and Zheng-An Wang and Si-Yun Zhou and Hao Li and Kui Zhao and Yue-Shan Xu and Wei-Guo Ma and Hao-Tian Liu and Cai-Ping Fang and Jia-Cheng Song and Tian-Ming Li and Jia-Chi Zhang and Yu Liu and Cheng-Lin Deng and Guangming Xue and Haifeng Yu and Kai Xu and Kaixuan Huang and Franco Nori and Heng Fan},
      year={2025},
      eprint={2508.20759},
      archivePrefix={arXiv},
      primaryClass={quant-ph},
      url={https://arxiv.org/abs/2508.20759}, 
}

@article{Mazzola2021,
  title = {Gauge-invariant quantum circuits for $U$(1) and Yang-Mills lattice gauge theories},
  author = {Mazzola, Giulia and Mathis, Simon V. and Mazzola, Guglielmo and Tavernelli, Ivano},
  journal = {Phys. Rev. Res.},
  volume = {3},
  issue = {4},
  pages = {043209},
  numpages = {15},
  year = {2021},
  month = {Dec},
  publisher = {American Physical Society},
  doi = {10.1103/PhysRevResearch.3.043209},
  url = {https://link.aps.org/doi/10.1103/PhysRevResearch.3.043209}
}

@article{Busnaina2025,
title = {Native three-body interactions in a superconducting lattice gauge quantum simulator},
  author = {Busnaina, J. H. and Shi, Z. and Alcaine-Cuervo, Jes\'us M. and Yang, Cindy X. and Nsanzineza, I. and Rico, E. and Wilson, C. M.},
  journal = {Phys. Rev. B},
  volume = {112},
  issue = {13},
  pages = {134514},
  numpages = {19},
  year = {2025},
  month = {Oct},
  publisher = {American Physical Society},
  doi = {10.1103/6prx-zmdz},
  url = {https://link.aps.org/doi/10.1103/6prx-zmdz}
}

@article{Popov2024,
  title = {Variational quantum simulation of U(1) lattice gauge theories with qudit systems},
  author = {Popov, Pavel P. and Meth, Michael and Lewestein, Maciej and Hauke, Philipp and Ringbauer, Martin and Zohar, Erez and Kasper, Valentin},
  journal = {Phys. Rev. Res.},
  volume = {6},
  issue = {1},
  pages = {013202},
  numpages = {19},
  year = {2024},
  month = {Feb},
  publisher = {American Physical Society},
  doi = {10.1103/PhysRevResearch.6.013202},
  url = {https://link.aps.org/doi/10.1103/PhysRevResearch.6.013202}
}

@misc{Jakobs2025,
      title={A Comprehensive Stress Test of Truncated Hilbert Space Bases against Green's function Monte Carlo in U(1) Lattice Gauge Theory}, 
      author={Timo Jakobs and Marco Garofalo and Tobias Hartung and Karl Jansen and Paul Ludwig and Johann Ostmeyer and Simone Romiti and Carsten Urbach},
      year={2025},
      eprint={2510.27611},
      archivePrefix={arXiv},
      primaryClass={hep-lat},
      url={https://arxiv.org/abs/2510.27611}, 
}

@misc{MirandaRiaza2025,
      title={Renormalized dual basis for scalable simulations of 2+1D compact quantum electrodynamics}, 
      author={Marc Miranda-Riaza and Pierpaolo Fontana and Alessio Celi},
      year={2025},
      eprint={2510.18594},
      archivePrefix={arXiv},
      primaryClass={quant-ph},
      url={https://arxiv.org/abs/2510.18594}, 
}

@article{Zhang2023,
  doi = {10.22331/q-2023-10-23-1148},
  url = {https://doi.org/10.22331/q-2023-10-23-1148},
  title = {Simulating gauge theories with variational quantum eigensolvers in superconducting microwave cavities},
  author = {Zhang, Jinglei and Ferguson, Ryan and K{\"{u}}hn, Stefan and Haase, Jan F. and Wilson, C.M. and Jansen, Karl and Muschik, Christine A.},
  journal = {{Quantum}},
  issn = {2521-327X},
  publisher = {{Verein zur F{\"{o}}rderung des Open Access Publizierens in den Quantenwissenschaften}},
  volume = {7},
  pages = {1148},
  month = oct,
  year = {2023}
}

@misc{Pardo2025,
doi = {10.1088/2058-9565/adce2a},
url = {https://doi.org/10.1088/2058-9565/adce2a},
year = {2025},
month = {apr},
publisher = {IOP Publishing},
volume = {10},
number = {3},
pages = {035011},
author = {Pardo, Guy and Bender, Julian and Katz, Nadav and Zohar, Erez},
title = {Truncation-free quantum simulation of pure-gauge compact QED using Josephson arrays},
journal = {Quantum Science and Technology}
}

@misc{Lin2025,
      title={Lattice field theory for superconducting circuits}, 
      author={Joshua Lin and Max Hays and Stephen Sorokanich III and Julian Bender and Phiala E. Shanahan and Neill C. Warrington},
      year={2025},
      eprint={2512.05851},
      archivePrefix={arXiv},
      primaryClass={quant-ph},
      url={https://arxiv.org/abs/2512.05851}, 
}

@misc{Stavenger2022,
      title={Bosonic Qiskit}, 
      author={Timothy J Stavenger and Eleanor Crane and Kevin Smith and Christopher T Kang and Steven M Girvin and Nathan Wiebe},
      year={2022},
      eprint={2209.11153},
      archivePrefix={arXiv},
      primaryClass={quant-ph},
      url={https://arxiv.org/abs/2209.11153}, 
}

@misc{Ale2025,
      title={Simulating quantum electrodynamics in 2+1 dimensions with qubits and qumodes}, 
      author={Victor Ale and Tommaso Rainaldi and Enrique Rico and Felix Ringer and George Siopsis},
      year={2025},
      eprint={2511.14506},
      archivePrefix={arXiv},
      primaryClass={quant-ph},
      url={https://arxiv.org/abs/2511.14506}, 
}

@article{PhysRevD.11.395,
  title = {Hamiltonian formulation of Wilson's lattice gauge theories},
  author = {Kogut, John and Susskind, Leonard},
  journal = {Phys. Rev. D},
  volume = {11},
  issue = {2},
  pages = {395--408},
  numpages = {0},
  year = {1975},
  month = {Jan},
  publisher = {American Physical Society},
  doi = {10.1103/PhysRevD.11.395},
  url = {https://link.aps.org/doi/10.1103/PhysRevD.11.395}
}

@phdthesis{Cottet2002,
  title={Implementation of a quantum bit in a superconducting circuit},
  author={Cottet, Audrey},
  year={2002},
  school={PhD Thesis, Universit{\'e} Paris 6}
}

@article{Didier2018,
  title = {Analytical modeling of parametrically modulated transmon qubits},
  author = {Didier, Nicolas and Sete, Eyob A. and da Silva, Marcus P. and Rigetti, Chad},
  journal = {Phys. Rev. A},
  volume = {97},
  issue = {2},
  pages = {022330},
  numpages = {13},
  year = {2018},
  month = {Feb},
  publisher = {American Physical Society},
  doi = {10.1103/PhysRevA.97.022330},
  url = {https://link.aps.org/doi/10.1103/PhysRevA.97.022330}
}

@article{Koch2007,
  title = {Charge-insensitive qubit design derived from the Cooper pair box},
  author = {Koch, Jens and Yu, Terri M. and Gambetta, Jay and Houck, A. A. and Schuster, D. I. and Majer, J. and Blais, Alexandre and Devoret, M. H. and Girvin, S. M. and Schoelkopf, R. J.},
  journal = {Phys. Rev. A},
  volume = {76},
  issue = {4},
  pages = {042319},
  numpages = {19},
  year = {2007},
  month = {Oct},
  publisher = {American Physical Society},
  doi = {10.1103/PhysRevA.76.042319},
  url = {https://link.aps.org/doi/10.1103/PhysRevA.76.042319}
}

@book{Olver2010,
  author    = {Olver, Frank W. J. and Lozier, Daniel W. and Boisvert, Ronald F. and Clark, Charles W.},
  title     = {The NIST Handbook of Mathematical Functions},
  publisher = {Cambridge University Press},
  year      = {2010},
  address   = {New York, NY},
}

@book{Arfken2013,
  author    = {George B. Arfken and Hans J. Weber and Frank E. Harris},
  title     = {Mathematical Methods for Physicists},
  edition   = {7},
  publisher = {Academic Press},
  year      = {2013},
  address   = {Boston},
  isbn      = {978-0-12-384654-9}
}

@article{Bauer2023,
  title = {Efficient representation for simulating {U}(1) gauge theories on digital quantum computers at all values of the coupling},
  author = {Bauer, Christian W. and Grabowska, Dorota M.},
  journal = {Phys. Rev. D},
  volume = {107},
  issue = {3},
  pages = {L031503},
  numpages = {6},
  year = {2023},
  month = {Feb},
  publisher = {American Physical Society},
  doi = {10.1103/PhysRevD.107.L031503},
  url = {https://link.aps.org/doi/10.1103/PhysRevD.107.L031503}
}

@article{Haase2021,
  doi = {10.22331/q-2021-02-04-393},
  url = {https://doi.org/10.22331/q-2021-02-04-393},
  title = {A resource efficient approach for quantum and classical simulations of gauge theories in particle physics},
  author = {Haase, Jan F. and Dellantonio, Luca and Celi, Alessio and Paulson, Danny and Kan, Angus and Jansen, Karl and Muschik, Christine A.},
  journal = {{Quantum}},
  issn = {2521-327X},
  publisher = {{Verein zur F{\"{o}}rderung des Open Access Publizierens in den Quantenwissenschaften}},
  volume = {5},
  pages = {393},
  month = feb,
  year = {2021}
}

@article{LevensonFalk2025,
  author  = {Eli M. Levenson-Falk and Sadman Ahmed Shanto},
  title   = {A Review of Design Concerns in Superconducting Quantum Circuits},
  journal = {Materials for Quantum Technology},
  volume  = {5},
  number  = {2},
  pages   = {022003},
  year    = {2025},
  doi     = {10.1088/2633-4356/ade10d}
}

@article{Nakamura1999,
  author  = {Nakamura, Y. and Pashkin, Yu. A. and Tsai, J. S.},
  title   = {Coherent control of macroscopic quantum states in a single-Cooper-pair box},
  journal = {Nature},
  volume  = {398},
  number  = {6730},
  pages   = {786--788},
  year    = {1999},
  doi     = {10.1038/19718}
}

@article{Martinis2002,
  author  = {J. M. Martinis and S. Nam and J. Aumentado and C. Urbina},
  title   = {Rabi Oscillations in a Large Josephson-Junction Qubit},
  journal = {Phys. Rev. Lett.},
  volume  = {89},
  pages   = {117901},
  year    = {2002},
  doi     = {10.1103/PhysRevLett.89.117901}
}

@article{Wendin2017,
  author = {G{\"o}ran Wendin},
  title = {Quantum Information Processing with Superconducting Circuits: A Review},
  journal = {Reports on Progress in Physics},
  volume = {80},
  number = {10},
  pages = {106001},
  year = {2017},
  doi = {10.1088/1361-6633/aa7e1a}
}

@article{Kwon2021,
  author = {S. Kwon and A. Tomonaga and G. Lakshmi Bhai and S. J. Devitt and J.‐S. Tsai},
  title = {Gate-based Superconducting Quantum Computing},
  journal = {Journal of Applied Physics},
  volume = {129},
  number = {4},
  pages = {041102},
  year = {2021},
  doi = {10.1063/5.0029735}
}

@article{Gao2021,
  author = {Y. Y. Gao and M. A. Rol and S. Touzard and C. Wang},
  title = {Practical Guide for Building Superconducting Quantum Devices},
  journal = {PRX Quantum},
  volume = {2},
  number = {4},
  pages = {040202},
  year = {2021},
  doi = {10.1103/PRXQuantum.2.040202}
}

@article{Rasmussen2021,
  author = {Simon Rasmussen and Kasper Christensen and Simon Pedersen and Lasse Kristensen and Thomas B{\ae}kkegaard and Niels Loft and Niels T. Zinner},
  title = {Superconducting Circuit Companion—An Introduction with Worked Examples},
  journal = {PRX Quantum},
  volume = {2},
  number = {4},
  pages = {040204},
  year = {2021},
  doi = {10.1103/PRXQuantum.2.040204}
}

@article{Pedersen1976,
    author = {Pedersen, N. F. and Soerensen, O. H. and Mygind, J. and Lindelof, P. E. and Levinsen, M. T. and Clark, T. D.},
    title = {Direct detection of the Josephson radiation emitted from superconducting thin‐film microbridges},
    journal = {Appl. Phys. Lett.},
    volume = {28},
    number = {9},
    pages = {562-564},
    year = {1976},
    month = {05},
    issn = {0003-6951},
    doi = {10.1063/1.88824},
    url = {https://doi.org/10.1063/1.88824}
}

@article{Soerensen1977,
    author = {Soerensen, O. H. and Mygind, J. and Pedersen, N. F. and Gubankov, V. N. and Levinsen, M. T. and Lindelof, P. E.},
    title = {Nonresonant detection of Josephson radiation from thin‐film microbridges},
    journal = {J. Appl. Phys.},
    volume = {48},
    number = {12},
    pages = {5372-5374},
    year = {1977},
    month = {12},
    issn = {0021-8979},
    doi = {10.1063/1.323542},
    url = {https://doi.org/10.1063/1.323542},
}

@article{karimi2024,
  title={Bolometric detection of Josephson radiation},
  author={Karimi, Bayan and Steffensen, Gorm Ole and Higginbotham, Andrew P and Marcus, Charles M and Levy Yeyati, Alfredo and Pekola, Jukka P},
  journal={Nat. Nanotechnol.},
  volume={19},
  number={11},
  pages={1613--1618},
  year={2024},
  publisher={Nature Publishing Group UK London},
  doi={10.1038/s41565-024-01770-7}
}

@book{auerbach2012interacting,
  title={Interacting electrons and quantum magnetism},
  author={Auerbach, Assa},
  year={2012},
  publisher={Springer Science \& Business Media}
}

@Article{kato1949perturbation,
  title         = {On the Convergence of the Perturbation Method. I},
  volume        = {4},
  issn          = {1347-4081},
  url           = {http://dx.doi.org/10.1143/ptp/4.4.514},
  doi           = {10.1143/ptp/4.4.514},
  number        = {4},
  journal       = {Progress of Theoretical Physics},
  publisher     = {Oxford University Press (OUP)},
  author        = {Kato, T.},
  year          = {1949},
  month         = dec,
  pages         = {514–523}
}

@book{kato2013perturbationbook,
  title={Perturbation theory for linear operators},
  author={Kato, Tosio},
  volume={132},
  year={2013},
  publisher={Springer Science \& Business Media}
}

@Article{takahashi1977half-filled,
  title         = {Half-filled Hubbard model at low temperature},
  volume        = {10},
  issn          = {0022-3719},
  url           = {http://dx.doi.org/10.1088/0022-3719/10/8/031},
  doi           = {10.1088/0022-3719/10/8/031},
  number        = {8},
  journal       = {Journal of Physics C: Solid State Physics},
  publisher     = {IOP Publishing},
  author        = {Takahashi, M},
  year          = {1977},
  month         = apr,
  pages         = {1289–7301}
}

@InBook{mila2010strong-coupling,
  title         = {Strong-Coupling Expansion and Effective Hamiltonians},
  isbn          = {9783642105890},
  issn          = {0171-1873},
  url           = {http://dx.doi.org/10.1007/978-3-642-10589-0_20},
  doi           = {10.1007/978-3-642-10589-0_20},
  booktitle     = {Introduction to Frustrated Magnetism},
  publisher     = {Springer Berlin Heidelberg},
  author        = {Mila, Frédéric and Schmidt, Kai Phillip},
  year          = {2010},
  month         = sep,
  pages         = {537–559}
}

@article{perez2006matrix,
author = {Perez-Garcia, D. and Verstraete, F. and Wolf, M. M. and Cirac, J. I.},
title = {Matrix product state representations},
year = {2007},
issue_date = {July 2007},
publisher = {Rinton Press, Incorporated},
address = {Paramus, NJ},
volume = {7},
number = {5},
issn = {1533-7146},
journal = {Quantum Info. Comput.},
month = jul,
pages = {401–430},
numpages = {30},
doi={10.5555/2011832.2011833},
}

@book{montangero2018introduction,
  title={Introduction to tensor network methods},
  author={Montangero, Simone and Montangero, Evenson and Evenson},
  year={2018},
  publisher={Springer}
}

@article{RevModPhys.93.045003,
  title = {Matrix product states and projected entangled pair states: Concepts, symmetries, theorems},
  author = {Cirac, J. Ignacio and P\'erez-Garc\'{\i}a, David and Schuch, Norbert and Verstraete, Frank},
  journal = {Rev. Mod. Phys.},
  volume = {93},
  issue = {4},
  pages = {045003},
  numpages = {65},
  year = {2021},
  month = {Dec},
  publisher = {American Physical Society},
  doi = {10.1103/RevModPhys.93.045003},
  url = {https://link.aps.org/doi/10.1103/RevModPhys.93.045003}
}

@article{weiss2019spectrum,
  title = {Spectrum and coherence properties of the current-mirror qubit},
  author = {Weiss, D. K. and Li, Andy C. Y. and Ferguson, D. G. and Koch, Jens},
  journal = {Phys. Rev. B},
  volume = {100},
  issue = {22},
  pages = {224507},
  numpages = {17},
  year = {2019},
  month = {Dec},
  publisher = {American Physical Society},
  doi = {10.1103/PhysRevB.100.224507},
  url = {https://link.aps.org/doi/10.1103/PhysRevB.100.224507}
}

@article{di2021efficient,
  title={Efficient modeling of superconducting quantum circuits with tensor networks},
  author={Di Paolo, Agustin and Baker, Thomas E and Foley, Alexandre and S{\'e}n{\'e}chal, David and Blais, Alexandre},
  journal={npj Quantum Inf.},
  volume={7},
  number={1},
  pages={11},
  year={2021},
  publisher={Nature Publishing Group UK London},
  doi={10.1038/s41534-020-00352-4}
}

@article{roy2023quantum,
  title = {Quantum electronic circuits for multicritical Ising models},
  author = {Roy, Ananda},
  journal = {Phys. Rev. B},
  volume = {108},
  issue = {23},
  pages = {235414},
  numpages = {9},
  year = {2023},
  month = {Dec},
  publisher = {American Physical Society},
  doi = {10.1103/PhysRevB.108.235414},
  url = {https://link.aps.org/doi/10.1103/PhysRevB.108.235414}
}

@article{rui2025tensor,
  title={Tensor-network representation of excitations in Josephson junction arrays},
  author={Rui, Emilio and Cohen, Joachim and Petrescu, Alexandru},
  journal={arXiv preprint arXiv:2510.08680},
  year={2025},
  url={https://arxiv.org/abs/2510.08680}, 
}

@Article{tenpy2024,
    title={{Tensor network Python (TeNPy) version 1}},
    author={Johannes Hauschild and Jakob Unfried and Sajant Anand and Bartholomew Andrews and Marcus Bintz and Umberto Borla and Stefan Divic and Markus Drescher and Jan Geiger and Martin Hefel and Kévin Hémery and Wilhelm Kadow and Jack Kemp and Nico Kirchner and Vincent S. Liu and Gunnar Möller and Daniel Parker and Michael Rader and Anton Romen and Samuel Scalet and Leon Schoonderwoerd and Maximilian Schulz and Tomohiro Soejima and Philipp Thoma and Yantao Wu and Philip Zechmann and Ludwig Zweng and Roger S. K. Mong and Michael P. Zaletel and Frank Pollmann},
    journal={SciPost Phys. Codebases},
    pages={41},
    year={2024},
    publisher={SciPost},
    doi={10.21468/SciPostPhysCodeb.41},
    url={https://scipost.org/10.21468/SciPostPhysCodeb.41},
}

\clearpage

\onecolumngrid

\appendix

\section{Circuit Quantization}
\label{app_sec:Circuit_quantization}

\subsection{Derivation of the U(1) minimal-coupling term}
\label{app_subsec:Node_flux_variables_3q}

\begin{figure}[b]
    \centering
    \includegraphics[scale=0.24]{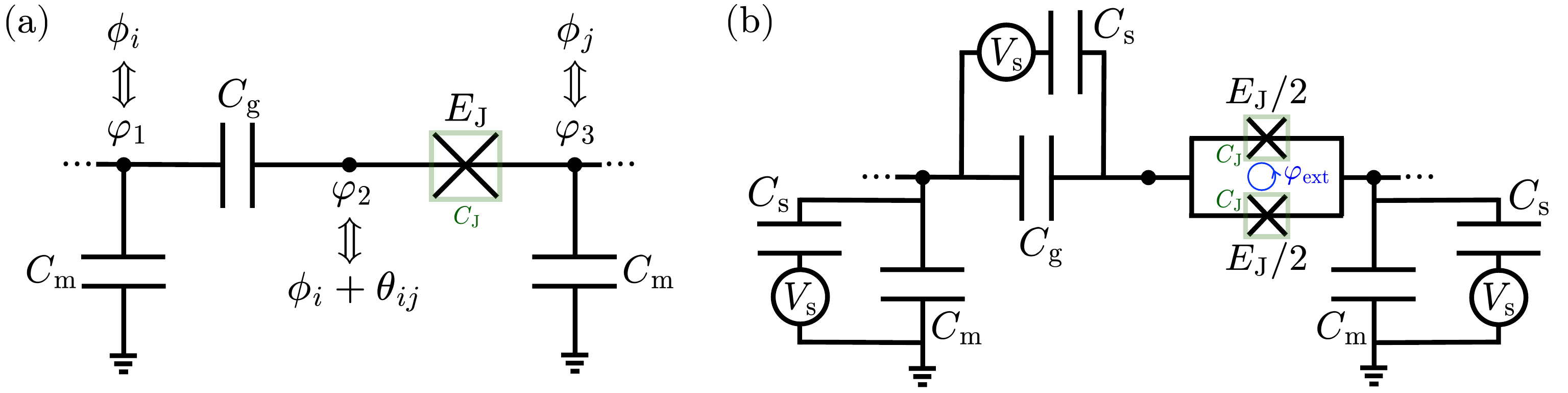}
    \caption{Lumped-element representation of the branch circuit. (a) Node-flux formulation realizing the minimal gauge-matter coupling $\cos{( \hat{\phi}_i+\hat{\theta}_{ij} - \hat{\phi}_j )}$. The circuit consists of three capacitors-two with capacitance $C_{\rm m}$ and one with capacitance $C_{\rm g}$-and a Josephson junction with Josephson energy $E_{\rm J}$. (b) Inclusion of classical control parameters to account for charge and flux offsets. Gate voltages $V_{\rm s}$ and gate capacitors with capacitance $C_{\rm s}$ introduce offset charges, while an external magnetic flux $\varphi_{\rm ext}$ threading a SQUID renders the Josephson energy tunable. In both panels, the green box indicates a small shunt capacitance $C_{\rm J}$ associated with the Josephson element.}
    \label{fig:Lumped_circuit_variables}
\end{figure}

We introduce dimensionless node phase (flux) variables $\varphi_i = 2\pi \Phi_0^{-1} \int_{-\infty}^{t} V_i(\tau)\, \mathrm{d}\tau $, where $V_i(\tau)$ denotes the time-dependent electrostatic potential at node $i$.
In terms of these variables, the Lagrangian of one side of the plaquette circuit reads~
\begin{equation}
    L = \frac{\Cm}{2} \left( \dot{\varphi}_1^2 + \dot{\varphi}_3^2 \right) + \frac{\Cg}{2} \left( \dot{\varphi}_3 - \dot{\varphi}_1 \right)^2 + \EJ \cos{\left( \varphi_3 - \varphi_2 \right)}.
\end{equation}
It is convenient to perform the linear transformation $\boldsymbol{\Phi} = \boldsymbol{\mathcal{A}}  \boldsymbol{\varphi}$, with $\boldsymbol{\varphi}^T = \left( \varphi_1 \ \ \varphi_2 \ \ \varphi_3 \right)$, $\boldsymbol{\Phi}^T = \left( \phi_i \ \ \theta_{ij} \ \ \phi_j \right)$ and
\begin{equation}
    \boldsymbol{\mathcal{A}} =
    \begin{pmatrix}
        1 & 0 & 0 \\
        -1 & 1 & 0 \\
        0 & 0 & 1
    \end{pmatrix}.
\end{equation}
In these variables, the Lagrangian becomes
\begin{equation}
    L_{ij} =  \frac{\Cm}{2} \left( \dot{\phi}_i^2 + \dot{\phi}_j^2 \right) + \frac{\Cg}{2} \dot{\theta}_{ij}^2 + \EJ \cos{\left( \phi_i + \theta_{ij} - \phi_j \right)},
\end{equation}
which explicitly realizes the desired gauge-matter interaction term [see Fig.~\ref{fig:Lumped_circuit_variables}(a)]. The new variables  $\phi$ and $\theta$ are naturally identified with matter (site) and gauge (link) d.o.f, respectively.
Since $\boldsymbol{\mathcal{A}}$ has full rank and a nonzero determinant, the transformation is invertible and does not introduce constraints.

Writing the Lagrangian in matrix form as $L_{ij} = \frac{1}{2} \boldsymbol{\dot{\Phi}}^T \boldsymbol{\mathcal{C}} \boldsymbol{\dot{\Phi}} + U(\boldsymbol{\Phi}) $, we define the conjugate charges $\boldsymbol{Q} = \partial L_{ij}/\partial \boldsymbol{\dot{\Phi}} $. 
Performing the Legendre transform, the Hamiltonian is $H_{ij} =\boldsymbol{Q}\cdot   \boldsymbol{\dot{\Phi}} (\boldsymbol{Q}) - L_{ij} = \frac{1}{2} \boldsymbol{Q}^T \boldsymbol{\mathcal{C}}^{-1} \boldsymbol{Q} - U(\boldsymbol{\Phi}) = 2e^2 \boldsymbol{n}^T \boldsymbol{\mathcal{C}}^{-1} \boldsymbol{n} - U(\boldsymbol{\Phi}) $, where $\boldsymbol{n} = \boldsymbol{Q}/(2e)$ is the charge-number operator. 
For the capacitance matrix $ \boldsymbol{\mathcal{C}} = {\rm diag}(C_{\rm m}, C_{\rm g}, C_{\rm m})$ the Hamiltonian becomes
\begin{equation}
    \begin{aligned}
        \hat{H}_{ij} &= m \, (\hat{n}_{\phi_i}^2 + \hat{n}_{\phi_j}^2  )
        + g \, \hat{n}_{\theta_{ij}}^2
        - \lambda \cos{(\hat{\phi}_i + \hat{\theta}_{ij} - \hat{\phi}_j)},
    \end{aligned}
    \label{eq:minimal_hamiltonian_appendix}
\end{equation}
where we have introduced the effective charging and Josephson energies $m \equiv 4 E_{\Cm} = 2e^2/\Cm$, $g \equiv 4 E_{\Cg}=2e^2/\Cg$, and $\lambda \equiv \EJ$. This yields the circuit Hamiltonian presented in the main text.

To assess the impact of charge and flux fluctuations on the circuit, we extend the setup by shunting each capacitor with a gate voltage source $V_{\rm s}$ through a capacitor with capacitance $C_{\rm s}$, and by replacing the Josephson junction with a SQUID threaded by an external magnetic flux $\varphi_{\rm ext}$ [see Fig.~\ref{fig:Lumped_circuit_variables}(b)]. 
Since $V_{\rm s}$ and $\varphi_{\rm ext}$ are externally controlled classical parameters, they are not quantized. 
The resulting Lagrangian, denoted $L_{ij} \equiv L_{ij} (C_{\rm s}, V_{\rm s},\varphi_{\rm ext})$, takes the form $L_{ij} = \frac{1}{2} \dot{\boldsymbol{\Phi}}^{T} \boldsymbol{\mathcal{C}} \dot{\boldsymbol{\Phi}} - {V}_{\rm s}  C_{\rm s}  \dot{\boldsymbol{\Phi}}+ U( {\boldsymbol{\Phi}})$, where $\dot{\boldsymbol{\Phi}}^T = (\dot{\phi}_i, \dot{\theta}_{ij}, \dot{\phi}_j)$. The capacitance matrix acquires diagonal corrections, $\boldsymbol{\mathcal{C}} = {\rm diag} (C_{\rm ms},C_{\rm gs}, C_{\rm ms})$, with $ C_{\rm ms} \equiv \Cm + C_{\rm s}, \, C_{\rm gs} \equiv \Cg + C_{\rm s} $. 
The conjugate charges are now $\boldsymbol{Q}   = \boldsymbol{\mathcal{C}}  \dot{\boldsymbol{\Phi}} - C_{\rm s} {V}_{\rm s} \boldsymbol{\mathbb{1}}$, which yields $ \dot{\boldsymbol{\Phi}} = \boldsymbol{\mathcal{C}}^{-1} \left( \boldsymbol{Q} + C_{\rm s} {V}_{\rm s}\boldsymbol{\mathbb{1}}\right)$. 
The Hamiltonian becomes $H_{ij} = \frac{1}{2} (\boldsymbol{Q}+C_{\rm s} {V}_{\rm s}\boldsymbol{\mathbb{1}})^T \boldsymbol{\mathcal{C}}^{-1} (\boldsymbol{Q}+C_{\rm s} {V}_{\rm s}\boldsymbol{\mathbb{1}} ) - \boldsymbol{\mathcal{U}}(\boldsymbol{\Phi}) = 2e^2 (\boldsymbol{n}-n_{\rm s} \boldsymbol{\mathbb{1}})^T \boldsymbol{\mathcal{C}}^{-1} (\boldsymbol{n}-n_{\rm s}  \boldsymbol{\mathbb{1}}) - U(\boldsymbol{\Phi}) $, where the offset charge is $n_{\rm s} = - C_{\rm s} {V}_{\rm s} / (2e)$. 
The resulting circuit Hamiltonian reads
\begin{equation}
    \begin{aligned}
        \hat{H}_{ij} (n_{\rm s}, \varphi_{\rm ext}) &= m_{\rm s} \left[ (\hat{n}_{\phi_i} - n_{\rm s})^2 + (\hat{n}_{\phi_j}- n_{\rm s})^2  \right]
        + g_{\rm s} (\hat{n}_{\theta_{ij}}- n_{\rm s})^2
        - \lambda (\varphi_{\rm ext})  \cos{(\hat{\phi}_i + \hat{\theta}_{ij} - \hat{\phi}_j)} ,
    \end{aligned}
    \label{eq:minimal_hamiltonian_appendix_2}
\end{equation}
with renormalized energy scales $m_{\rm s} \equiv  2e^2/C_{\rm ms}$, $g_{\rm s}\equiv 2e^2/C_{\rm gs}$, and a flux-tunable Josephson coupling $ \lambda(\varphi_{\rm ext})\equiv E_{\rm J}\cos{ (\varphi_{\rm ext}/2) } $. 
Throughout, constants modified by the gate capacitance $C_{\rm s}$ are denoted by the subscript ``s".

\subsection{Effects of Josephson junction shunt capacitances}
\label{app_subsec:JJ_with_CJ_3q}

Usually, Josephson junctions have an intrinsic shunt capacitance $\CJ$ [see Fig. \ref{fig:Lumped_circuit_variables}(a)]. 
Including this contribution, the branch circuit Lagrangian becomes
\begin{equation}
    \begin{aligned}
        L_{ij}' =& \frac{\Cm}{2} (\dot{\phi}_i^2 + \dot{\phi}_j^2) + \frac{\Cg}{2} \dot{\theta}^2_{ij} + \frac{\CJ}{2} (\dot{\phi_i} + \dot{\theta}_{ij} - \dot{\phi}_j)^2  + E_{\rm J} \cos{(\phi_i + {\theta}_{ij} - \phi_j)}
    \end{aligned}
\end{equation}
The corresponding capacitance matrix is no longer diagonal and reads
\begin{equation}
    \boldsymbol{\mathcal{C}}' =
    \begin{bmatrix}
        C_{\rm m} + C_{\rm J} & C_{\rm J} & - C_{\rm J} \\
        C_{\rm J} & C_{\rm g} + C_{\rm J} & - C_{\rm J} \\
        - C_{\rm J} & - C_{\rm J} & C_{\rm m} + C_{\rm J}
    \end{bmatrix}.
\end{equation}
The Hamiltonian is given by $H'_{ij} = \frac{1}{2}  \boldsymbol{Q}'^{T}\cdot \boldsymbol{\mathcal{C}}'^{-1} \boldsymbol{Q}' - E_{\rm J} \cos{(\phi_i + {\theta}_{ij} - \phi_j)} )$, where $\boldsymbol{Q}'$ denotes the vector of conjugate charges. 
The inverse capacitance matrix is
\begin{equation}
    \begin{aligned}
        \boldsymbol{\mathcal{C}}'^{-1} &= \frac{1}{D}
        \begin{bmatrix}
            C_{\rm g} C_{\rm m} + C_{\rm J} (C_{\rm g} + C_{\rm m}) & - C_{\rm J} C_{\rm m} & C_{\rm J} C_{\rm g}\\
            - C_{\rm J} C_{\rm m} & C_{\rm m}^2 + 2 C_{\rm J} C_{\rm m} & C_{\rm J} C_{\rm m} \\
            C_{\rm J} C_{\rm g} & C_{\rm J} C_{\rm m} & C_{\rm g} C_{\rm m} + C_{\rm J} (C_{\rm g} + C_{\rm m})
        \end{bmatrix},
    \end{aligned}
\end{equation}
with determinant $D \equiv \det \boldsymbol{\mathcal{C}}' = C_{\rm m}^2 (C_{\rm g} + C_{\rm m}) + 2 C_{\rm m} C_{\rm J} C_{\rm g}$. 
Assuming $\CJ \ll \Cm, \Cg$, the inverse capacitance matrix can be expanded perturbatively. 
Retaining terms up to first order in $C_{\rm J}$, the Hamiltonian becomes
\begin{equation}
    \begin{aligned}
        \hat{H}_{ij}' \simeq& \ \ m' (\hat{n}_{\phi_i}^2 +  \hat{n}_{\phi_j}^2 ) + g' \hat{n}_{\theta_{ij}}^2
        - \lambda \cos{(\hat{\phi}_i + \hat{\theta}_{ij} - \hat{\phi}_j)} + \chi ( \hat{n}_{\theta_{ij}} \hat{n}_{\phi_j} - \hat{n}_{\phi_i} \hat{n}_{\theta_{ij}})  + \eta \hat{n}_{\phi_i} \hat{n}_{\phi_j},
    \end{aligned}
\end{equation}
where $m' \equiv  2 e^2 (C_{\rm m } - C_{\rm J})/C_{\rm m}^2$, $g' \equiv 2 e^2 (C_{\rm g} - C_{\rm J})/C_{\rm g}^2$, $\chi \equiv 4 e^2 C_{\rm J}/(C_{\rm g}C_{\rm m})$, $\eta \equiv 4 e^2 C_{\rm J}/C_{\rm m}^2$, and $\lambda \equiv E_{\rm J}$. 
Thus, the Josephson shunt capacitance renormalizes the charging energies and introduces small gauge-matter and matter-matter cross-coupling terms, while preserving gauge invariance.

The same procedure applies when in the presence of classical offset charges $n_{\rm s}$ and a flux-tunable Josephson energy $\lambda (\varphi_{\rm ext}) \equiv E_{\rm J} \cos{(\varphi_{\rm ext}/2 )}$ [see Fig.~\ref{fig:Lumped_circuit_variables}(b)], yielding
\begin{equation}
    \begin{aligned}
        \hat{H}' (n_{\rm s}, \varphi_{\rm ext}) =& m'_{\rm s} \left[ (\hat{n}_{\phi_i} - n_{\rm s})^2 + (\hat{n}_{\phi_j}- n_{\rm s})^2  \right]
        + g'_{\rm s} (\hat{n}_{\theta_{ij}}- n_{\rm s})^2
        - \lambda (\varphi_{\rm ext})  \cos{(\hat{\phi}_i + \hat{\theta}_{ij} - \hat{\phi}_j)} \\
        &+ \chi_{\rm s} \left[ (\hat{n}_{\theta}- n_{\rm s}) (\hat{n}_{\phi'}- n_{\rm s}) - (\hat{n}_{\phi}- n_{\rm s}) (\hat{n}_{\theta}- n_{\rm s}) \right]  + \eta_{\rm s} (\hat{n}_{\phi}- n_{\rm s}) (\hat{n}_{\phi'}- n_{\rm s}),
    \end{aligned}
\end{equation}
where all constants are appropriately renormalized to account for both $C_{\rm J}$ and the gate capacitance $C_{\rm s}$, i.e., $m'\rightarrow m'_{\rm s}$, $g'\rightarrow g'_{\rm s}$, $\chi \rightarrow \chi_{\rm s}$ and $\eta \rightarrow \eta_{\rm s}$.

\section{Matching Gauss's laws with conserved quantities of the circuit. Physical degrees of freedom.}
\label{app:Gauss_match_dof}

\begin{figure}[t]
    \centering
    \includegraphics[scale=0.26]{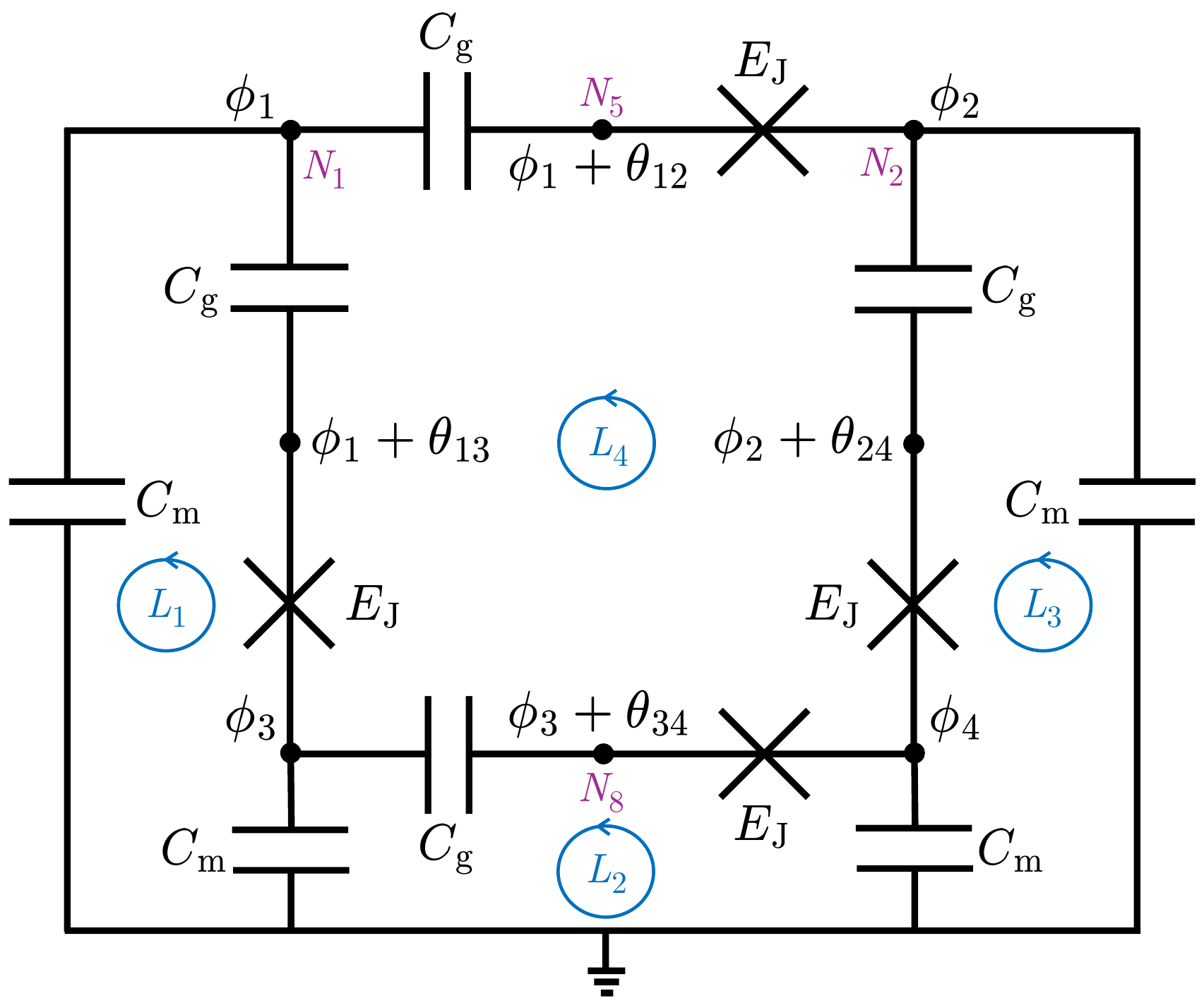}
    \caption{Lumped-element representation of the plaquette circuit, indicating the independent nodes and loops as determined by Kirchhoff’s current law (KCL) and Kirchhoff’s voltage law (KVL). The system has four physical degrees of freedom.}
    \label{fig:Kirchoff_plaquette}
\end{figure}

We now analyze the plaquette circuit to clarify its structure as a constrained system, in which the number of conserved quantities follows directly from Kirchhoff’s laws. 

As shown in Fig.~\ref{fig:Kirchoff_plaquette}, the circuit contains eight capacitive elements and four Josephson junctions, yielding $B=12$ branches and $N=9$ nodes. While branch fluxes and charges provide a convenient canonical description, they are not independent due to Kirchoff's laws. Kirchhoff’s current law (KCL) yields $N-1 = 8$ independent constraints due to global charge conservation, while Kirchhoff’s voltage law (KVL) provides $L = B - (N-1) = 4$ independent loop constraints, in accordance with the fundamental theorem of network topology. As a result, only four independent dynamical d.o.f remain. These correspond to the physical d.o.f of the plaquette, while the remaining variables generate conserved charges that implement Gauss's laws.

To formalize this, consider a general Hamiltonian of the form
\begin{equation}
    H= \frac{1}{2} \boldsymbol{n}^T \mathbf{K}^{-1} \boldsymbol{n}+\sum_{\alpha=1}^{D}V_\alpha(\mathbf{v}_\alpha\cdot\boldsymbol{\varphi}),
\end{equation}
with canonical variables $\varphi_i$ and $n_{\varphi_i}$ satisfying $[ \hat{\varphi}_i, \hat{n}_{\varphi_j}] = i\delta_{ij}$, $i\in\{1,\ldots,N\}$.
While the kinetic term depends on all $N$ charge variables, the potential depends only on $D<N$ independent linear combinations of fluxes.
The vectors $\mathbf{v}_\alpha$ span a $D$-dimensional subspace $V_{D} \subset \mathbb{R}^{N}$, leaving a $(N-D)$-dimensional nullspace 
\begin{equation}
    V_{N-D} = \{ \mathbf{u}\in \mathbb{R} ^N; \mathbf{u}\cdot \mathbf{v}_{\alpha} = 0\ \forall \alpha\},
\end{equation}
corresponding to ``flat" directions of the potential. Any combination of fluxes in this nullspace does not appear in the Hamiltonian and therefore represents a conserved quantity.

An orthogonal change of basis $\boldsymbol{O}$, with $\boldsymbol{O}\mathbf{v}_\alpha= (0_{N-D},\mathbf{w}_\alpha)^T$, isolates these directions: $\boldsymbol{\varphi}'=\boldsymbol{O}\boldsymbol{\varphi}=(\boldsymbol{\varphi}_{\mathrm{ign}},\boldsymbol{\varphi}_{\mathrm{v}})^T$, $\boldsymbol{n}'=\boldsymbol{O}\boldsymbol{n}=(\boldsymbol{n}_{\mathrm{ign}},\boldsymbol{n}_{\mathrm{v}})^T$. 
The variables $\boldsymbol{\varphi}_{\mathrm{ign}}$ are ignorable coordinates, and their conjugate charges $\boldsymbol{n}_{\mathrm{ign}}$ commute with the Hamiltonian, $[\hat{H}_{\rm P}, \hat{n}_{\mathrm{ign},i}]=0$, signaling continuous symmetries.

For the plaquette, we take $\boldsymbol{\varphi}^T = (\phi_1,\phi_2,\phi_3,\phi_4,\theta_{12},\theta_{24},\theta_{13},\theta_{34})$, $\boldsymbol{n}^T =(n_1,n_2,n_3,n_4,n_{12},n_{24},n_{13},n_{34})$, with $\mathbf{K}^{-1} = {\rm diag}(m,m,m,m,g,g,g,g)$.
The potential terms are defined by
\begin{equation}
    \mathbf{v}_\alpha\cdot\boldsymbol{\varphi} =
    \begin{pmatrix}
        1 & -1 & 0 & 0 & 1 & 0 & 0 & 0 \\
        0 & 1 & 0 & -1 & 0 & 1 & 0 & 0 \\
        1 & 0 & -1 & 0 & 0 & 0 & 1 & 0 \\
        0 & 0 & 1 & -1 & 0 & 0 & 0 & 1
    \end{pmatrix} \cdot \boldsymbol{\varphi} =
    \begin{pmatrix}
        \phi_1 + \theta_{12} - \phi_2 \\
        \phi_2 + \theta_{24} - \phi_4 \\
        \phi_1 + \theta_{13} - \phi_3 \\
        \phi_3 + \theta_{34} - \phi_4
    \end{pmatrix},
\end{equation}
for $\alpha\in \{ 12,24,13,34 \}$. 
These vectors have rank 4, leaving a four-dimensional nullspace. 
Solving $\mathbf{u}\cdot \mathbf{v}_{\alpha} = 0$ yields a basis of ignorable flux combinations. 
A convenient one is
\begin{equation}
    \begin{split}
        \boldsymbol{\varphi}_{\mathrm{ign}} = \mathbf{u}\cdot \boldsymbol{\varphi} =
        \begin{pmatrix}
            1 & 0 & 0 & 0 & -1 & 0 & -1 & 0 \\
            0 & 1 & 0 & 0 & 1 & -1 & 0 & 0 \\
            0 & 0 & 1 & 0 & 0 & 0 & 1 & -1 \\
            0 & 0 & 0 & 1 & 0 & 1 & 0 & 1
        \end{pmatrix} \cdot \boldsymbol{\varphi} =
        \begin{pmatrix}
            \phi_1 - ( \theta_{12} + \theta_{13}) \\
            \phi_2 - ( \theta_{24} - \theta_{12}) \\
            \phi_3 - ( \theta_{34} - \theta_{13}) \\
            \phi_4 + ( \theta_{34} + \theta_{24})
        \end{pmatrix}
    \end{split}.
\end{equation}
The corresponding conjugate charges (up to normalization) are
\begin{equation}
    \begin{split}
        \boldsymbol{n}_{\mathrm{ign}} = \mathbf{u}\cdot \boldsymbol{n}=
        \begin{pmatrix}
            n_1 - ( n_{12} + n_{13}) \\
            n_2 - ( n_{24} - n_{12}) \\
            n_3 - ( n_{34} - n_{13}) \\
            n_4 + ( n_{34} + n_{24})
        \end{pmatrix}
    \end{split}.
\end{equation}
These operators coincide with the Gauss-law generators defined in Sec.~\ref{sec:sc_circuit}, $\boldsymbol{G} \equiv \boldsymbol{n}_{\rm ign}$.
They are conserved quantities, $[H_{\rm P}, G_i]=0$, and physical states satisfy $\hat{G}_i \ket{\Psi} = q_i \ket{\Psi}$, with $q_i$ denoting static charges. Gauge fixing allows restriction to the zero-charge sector, $\hat{G}_i|\psi\rangle=0$, which reduces the Hilbert space dimension from $N^8 \rightarrow N^4$. 
A convenient parametrization selects the gauge charges $n_{ij}$ as independent variables, expressing the matter charges as
\begin{equation}
    n_1 = n_{12} + n_{13}, \quad
    n_2 = n_{24} - n_{12}, \quad
    n_3 = n_{34} - n_{13}, \quad
    n_4 = - n_{24} - n_{34}.
    \label{eq:resolved_matter_sites}
\end{equation}
This choice respects all constraints and highlights the four independent d.o.f of the plaquette.

\section{Schr\"odinger equation and Mathieu functions}
\label{app_sec:Mathieu}

The plaquette Hamiltonian \eqref{eq:plaquette_hamiltonian} can be decomposed into a kinetic and a potential contribution,
\begin{equation}
    \begin{split}
        \hat{K}(m, g)    & \equiv m\sum_{i} \hat{n}_i^2 + g \sum_{\link} \hat{n}_{ij}^2, \\
        \hat{V}(\lambda) & \equiv - \lambda \sum_{\link} \cos(\hat{\phi}_i + \hat{\theta}_{ij} - \hat{\phi}_j),
        \label{eq:plaq_hamiltonian_divided}
    \end{split}
\end{equation}
such that $\hat{H}_{\text{P}}(m,g,\lambda) = \hat{K}(m, g) + \hat{V}(\lambda)$.

As discussed in Appendix~\ref{app:Gauss_match_dof}, imposing Gauss’s laws allows one to express the kinetic term solely in terms of the gauge charge operators $\hat{n}_{ij}$, leaving the matter charges $\hat{n}_i$ non-dynamical and fully determined by the gauge d.o.f \eqref{eq:resolved_matter_sites}. 
The kinetic term then reads
\begin{equation}
    \hat{K}(m, g)
    = (g + 2m)\left( \hat{n}_{12}^2 + \hat{n}_{24}^2 + \hat{n}_{13}^2 + \hat{n}_{34}^2 \right)
    + 2m \left(\hat{n}_{12} \hat{n}_{13} + \hat{n}_{24} \hat{n}_{34} - \hat{n}_{24} \hat{n}_{12} - \hat{n}_{34} \hat{n}_{13} \right),
\end{equation}
which can be written in matrix form as
\begin{equation}
    \hat{K}(m, g) =
    \begin{pmatrix}
        \hat{n}_{12} & \hat{n}_{24} & \hat{n}_{13} & \hat{n}_{34}
    \end{pmatrix}
    \begin{pmatrix}
        g + 2m & -m      & m   &    \\
        -m      & g + 2m &      & m \\
        m     &        & g+2m & -m  \\
        & m     & -m    & g+2m
    \end{pmatrix}
    \begin{pmatrix}
        \hat{n}_{12} \\ \hat{n}_{24} \\ \hat{n}_{13} \\ \hat{n}_{34}
    \end{pmatrix}
    \label{eq:gauge_kinetic_term_matrix_form}
\end{equation}
Isolating the dependency on $m$ and $g$ in $\hat{K}$, we can write $\hat{K}(m, g) = g \hat{\bm{n}}^{T} \hat{\bm{n}} + m \hat{\bm{n}}^T M \hat{\bm{n}}$, where $\hat{\bm{n}} = (\hat{n}_{12},\, \hat{n}_{24},\, \hat{n}_{13},\, \hat{n}_{34})^T$.
The matrix $M$ determines the main kinetic features of the plaquette Hamiltonian.
From \eqref{eq:gauge_kinetic_term_matrix_form} we see that is
\begin{equation}
    M =
    \begin{pmatrix}
        2  & -1 &  1 &    \\
        -1 &  2 &    &  1 \\
        1  &    &  2 & -1 \\
           &  1 & -1 &  2
    \end{pmatrix}.
\end{equation}
Its eigenvectors $\bm{v}_i$ and corresponding eigenvalues $\lambda_i$ are
\begin{equation}
    \begin{cases}
        \bm{v}_1 = (1,\, -1,\,  1,\, -1)^T & \lambda_1 = 4, \\
        \bm{v}_2 = (1,\,  0,\,  0,\, 1)^T  & \lambda_2 = 2, \\
        \bm{v}_3 = (0,\,  1,\, 1,\, 0)^T   & \lambda_3 = 2, \\
        \bm{v}_4 = (1,\,  1,\, -1,\, -1)^T & \lambda_4 = 0.
    \end{cases}
\end{equation}
The lowest eigenvector corresponds to the linear combination $\bm{v}_4 \cdot \hat{\bm{n}} = (\hat{n}_{12} + \hat{n}_{24} - \hat{n}_{13} - \hat{n}_{34})$, which play an important role in the limit $m \to \infty$.
In fact, it corresponds to the single effective degree of freedom of the plaquette that emerges at low energy in the limit $m \to \infty$.
Concretely, this single degree of freedom corresponds to the total flux $\hat{\theta}_{\square} = (\hat{\theta}_{12} + \hat{\theta}_{24} - \hat{\theta}_{34} - \hat{\theta}_{13})$ going around the plaquette and a state with non-zero value of $\hat{\theta}_{\square}$ can be interpreted as a vortex state.
Furthermore, when $m \to \infty$, the low energy spectrum is determined solely by the fluctuations in $\hat{\theta}_{\square}$, because fluctuations in the matter charges are forbidden.
Given that we have identified $\hat{\theta}_{\square}$ as the vortex degree of freedom of the plaquette, this justifies the definition of the displacement operator $U_{\text{vortex}} \equiv e^{i \Theta (\hat{n}_{12} + \hat{n}_{24} - \hat{n}_{34} - \hat{n}_{13})/4}$ as the vortex creation operator.

Since the matter sites are no longer dynamical, their phases $\hat{\phi}_i$ can be eliminated from the potential term, which reduces to
\begin{equation}
    \hat{V}(\lambda) = -\lambda \sum_{\link}  \cos(\hat{\theta}_{ij}).
\end{equation}
With this choice of physical basis, we see that in the case $m = 0$ the system is equivalent to four decoupled gauge links.
Turning on the mass term, i.e.~$m \neq 0$, now means introducing a perturbation in the kinetic term, as it can be seen in \eqref{eq:gauge_kinetic_term_matrix_form}, that introduces coupling between the different gauge links.
Even when turning on the mass term, the potential remains diagonal in the phase basis of the gauge links.
We will later show how to solve the case $m = 0$ (\emph{the decoupled limit}) by putting the Schr\"odinger equation of the system in the phase basis in a Mathieu equation form.

Another limit that is worthwhile to look at is the case where $m \gg \lambda, g$, which we call the \emph{static matter limit}.
When $m \gg 1$, the cost of introducing matter charges, i.e., having $n_i \neq 0$, has a high energy cost compared to the other terms.
Therefore, in the low-energy sector, we can effectively take $n_i = 0$ for all the matter sites.
Given \eqref{eq:resolved_matter_sites}, this leads to
\begin{equation}
    n_{12} = n_{24} = - n_{13} = - n_{34}
    \qquad \text{when $m \to \infty$},
\end{equation}
which means that there is only one effective degree of freedom.
Physically, this can be understood in the following way.
Imposing $n_i = 0$ for all $i = 1, \dots, 4$ removes the dynamical matter d.o.f from the system, which means that gauge constraints only actually affect the gauge link d.o.f.
Moreover, the four gauge constraints are not all independent because global charge conservation has to be taken into account.
Therefore, with three independent constraints and four degrees of freedom, we are left with only one effective degree of freedom.

It is interesting to compare this picture with the decoupled limit.
In the latter case, when $m \to 0$, the cost of introducing matter charges vanishes.
Therefore, given any configuration of gauge link charges $n_{ij}$, the necessary matter charges $n_i$ that are needed to satisfy the gauge constraints can be introduced with essentially no cost in energy.
As a consequence, the gauge links can behave independently of each other.

\subsection{Decoupled limit and Mathieu equation}
\label{sub:decoupled_limit_and_mathieu_equation}

Consider now the decoupled limit $m = 0$, where the Hamiltonian is in the form
\begin{equation}
    \hat{H}(m = 0, g, \lambda) = \sum_{\link} \left( g \hat{n}_{ij}^2 - \lambda \cos \hat{\theta}_{ij} \right )
    \label{eq:hamiltonian_decoupled_limit}
\end{equation}
As we can see, the Hamiltonian is just a sum of single-body terms $\hat{H}_{ij} = g \hat{n}_{ij}^2 - \lambda \cos \hat{\theta}_{ij}$, each acting on a different gauge link; each gauge link can be solved independently.

We now move on to solving a single gauge link.
In the phase basis $\{\ket{\theta} : \theta \in [-\pi, \pi)\}$, the charge operator $\hat{n}$ acts as a derivative in the $\theta$-space, that is, $\hat{n} \equiv -i \partial_{\theta}$.
Therefore, the Schr\"odinger equation $\hat{H} \Psi(\theta) = E \Psi(\theta)$ is recast into
\begin{equation}
    \left[
        \dv[2]{}{\theta} + \left( \frac{E}{g} + \frac{\lambda}{g} \cos \theta  \right)
        \right] \Psi(\theta) = 0.
    \label{eq:schroedinger_single_link}
\end{equation}
If we define a new variable $x$ such that $\theta = 2x$, we have
\begin{equation}
    \left[
        \dv[2]{}{x} + \left( \frac{4 E}{g} + \frac{4 \lambda}{g} \cos(2 x) \right)
        \right] \Psi(x) = 0
    \label{eq:schroedinger_single_link_mathieu}
\end{equation}
and we recover the Mathieu equation
\begin{equation}
    \dv[2]{y}{x} + (a - 2q \cos(2x))y = 0
    \label{eq:mathieu_equation}
\end{equation}
with $q = - 2 \lambda / g$ and characteristic number $a = 4 E / g$.
The solutions to Schr\"odinger's equation \eqref{eq:schroedinger_single_link} are required to be periodic in $\theta$ with period $2 \pi$, which means that the solutions to Mathieu's equation \eqref{eq:schroedinger_single_link_mathieu} have to be periodic in $x$ with period $\pi$.
Therefore, the eigenfunctions are the $\pi$-periodic cosine-elliptic $\cen[2n](q; x)$ and sine-elliptic $\sen[2n](q; x)$ and the eigenvalues are given by the series $a_{2n}(q)$ and $b_{2n}(q)$ of their respective characteristic numbers.
We now analyze two different limits of the single link Hamiltonian: the harmonic oscillator limit for $q \gg 1$ and the free particle limit $q \to 0$.

\section{Effective Hamiltonian in the static matter regime}
\label{app_sec:Heff_static_m}

In this appendix, we develop the perturbative analysis of the circuit in the regime where $m \gg \lambda, g$ and show how the plaquette term arises by computing the effective Hamiltonian up to the fourth order.

First, we give a physical intuition on how this result is possible.
In the regime $m \gg \lambda, g$, which we call the \emph{static matter regime},
the total Hilbert space of the system can be partitioned into two sectors $\mathcal{H}_0$ and $\overline{\mathcal{H}}_0$: one where $n_i = 0$ for all sites ($\mathcal{H}_0$) and the other where there are sites with $n_i \neq 0$ ($\overline{\mathcal{H}}_0$).
The states in the first sector can be assumed to have a lower energy, with respect to the other sector, and to also contain the ground state.
This is true at least when considering relatively low occupation numbers $n_{ij}$
\footnote{
    Technically speaking, given that the eigenvalues $n_{ij}$ are unbounded, highly excited states in $\mathcal{H}_0$ can have higher energy than some states in $\overline{\mathcal{H}}_0$ if the $n_{ij}$ are large enough. Therefore, true partitioning of the full Hilbert space is not possible. For this reason, we assume a low energy regime where the $n_{ij}$ are relatively low, to morally achieve a separation between the two sectors.
}.
Given the energy gap between the two sectors, the first excited states above the ground state can also be assumed to be in $\mathcal{H}_0$.
This is possible if the effective potential for the gauge flux variables $\hat{\theta}_{ij}$ is of the form $\cosine(\hat{\theta}_{12} + \hat{\theta}_{24} -\hat{\theta}_{34} - \hat{\theta}_{13})$, which does not involve matter d.o.f.
This can be realized by allowing a low-energy state $\ket{\varphi} \in \mathcal{H}_0$ to go through three ``virtual'' states, i.e. $\ket{\zeta_i} \notin \mathcal{H}_0$, in order to reach another state $\ket{\varphi^{\prime}} \in \mathcal{H}_0$, with $\Delta n_{ij} = \pm 1$.
This process is analogous to the superexchange in the half-filled Hubbard model in the large-$U$ limit~\cite{auerbach2012interacting}.

\subsection{Perturbation theory review}
\label{sub:perturbation_theory_review}

In order to develop a perturbative analysis of the circuit, we first quickly review the resolvent approach, first developed by Kato \cite{kato1949perturbation,kato2013perturbationbook}, which allow to obtain an effective Hamiltonian expressed purely through the unperturbed eigenfunctions.
In the following, we use the formulation of Takahashi \cite{takahashi1977half-filled, mila2010strong-coupling}.

Assume the Hamiltonian is in the form
\begin{equation}
    \hat{H} = \hat{H}_0 + \lambda \hat{V}
\end{equation}
where $\hat{H}_0$ is the unperturbated (or zeroth order) Hamiltonian and $\hat{V}$ is the perturbation with coupling $\lambda$.
Assume that $\hat{H}_0$ has an $m$-fold degenerate level $E_0$, which in our case corresponds to the ground state.
We call $\mathcal{H}_0$ the subspace spanned by the \emph{unpertubated states} at level $E_0$ and let $\Proj_0$ be the projector onto $\mathcal{H}_0$: $\Proj_0^2 = \Proj_0^{\dagger} = \Proj_0$ and $\hat{H}_0 \Proj_0 = \Proj_0 \hat{H}_0 = E_0 \Proj_0$.
Now, let $\mathcal{H}$ be the space spanned by the \emph{perturbated states} from $\mathcal{H}_0$ and $\Proj$ the projector onto $\mathcal{H}$.
Similarly $ \Proj^2 = \Proj^{\dagger} = \Proj $ and $ \hat{H} \Proj = \Proj \hat{H} $.

We can express $\Proj$ via an integral identity using the \emph{resolvent} $R(z) = (z - \hat{H})^{-1}$ of $\hat{H}$:
\begin{equation}
    \Proj
    = \frac{1}{2 \pi i} \oint_{C} \dd z R(z)
    \label{eq:integral_identity_proj}
\end{equation}
where $C$ is a contour in the complex plane that contains no spectrum of $\hat{H}_0$ \emph{except} for $E_0$.
The resolvent $R(z)$ can now be expressed in terms of the resolvent $R_0(z) = (z - \hat{H}_0)^{-1}$ of the unperturbated Hamiltonian $\hat{H}_0$ and the perturbation $\hat{V}$:
\begin{equation}
    R(z) = R_0 (z) \sum_{N = 0}^{\infty} \lambda^N (\hat{V} R_0(z))^N.
    \label{eq:resolvent_expansion}
\end{equation}
This is possible thanks to the series expansion $(X - Y)^{-1} = X^{-1} \sum_{N = 0}^{\infty} (Y X^{-1})^N$.
Furthermore, the resolvent $R_0(z)$ can be further decomposed into
\begin{equation}
    R_0(z) = \frac{\Proj_0}{(z - E_0)}  + S(z)
    \quad \text{with} \quad
    S(z) = \frac{\Id - \Proj_0}{(z - \hat{H}_0)},
    \label{eq:reduced_resolvent}
\end{equation}
where the latter is also known as the reduced resolvent.
As shown in \cite{kato1949perturbation}, inserting \eqref{eq:resolvent_expansion} into \eqref{eq:integral_identity_proj} and using \eqref{eq:reduced_resolvent} yields
\begin{equation}
    \Proj = \Proj_0 -
    \sum_{N = 1}^{\infty} \lambda^N \sum_{k_1 + \dots + k_{N+1} = N, \, k_i \geq 0}
    \Sproj^{k_1} \hat{V} \Sproj^{k_2} \hat{V} \cdots \hat{V} \Sproj^{k_{N+1}},
    \label{eq:projector_expansion}
\end{equation}
where
\begin{equation}
    \Sproj^0 \equiv - \Proj_0, \qquad
    \Sproj^k = S(z = E_0)^k = \qty(\frac{\Id - \Proj_0}{E_0 - \hat{H}_0})^k
    \quad \text{for $k \geq 1$}.
    \label{eq:expansion_resolvent_terms}
\end{equation}

Following \cite{takahashi1977half-filled}, we consider the operator $\Gamma$ that maps an unperturbated vector $\ket{\varphi} \in \mathcal{H}_0$ into a perturbated one $\ket{\psi} \in \mathcal{H}$, i.e.~$\ket{\psi} = \Gamma \ket{\varphi}$.
It is defined as
\begin{equation}
    \Gamma \equiv \Proj \Proj_0 (\Proj_0 \Proj \Proj_0)^{-1/2}
    \quad \text{with} \quad
    (\Proj_0 \Proj \Proj_0)^{-1/2} \equiv \Proj_0 + \sum_{k = 1}^{\infty} \frac{(2k - 1)!!}{(2k)!!} \qty[\Proj_0 (\Proj_0 - \Proj) \Proj_0]^{k}.
    \label{eq:wave_operator}
\end{equation}
Then, the effective Hamiltonian is simply defined as
\begin{equation}
    \hat{H}_{\text{eff}} = \Gamma^{\dagger} \hat{H} \Gamma,
\end{equation}
which allow us to replace the eigenvalue problem $(\hat{H} - E) \ket{\psi} = 0$ in the full Hilbert space with $(\hat{H}_{\text{eff}} - E) \ket{\phi} = 0$ in the reduced space $\mathcal{H}_0$.
Given \eqref{eq:projector_expansion} and \eqref{eq:wave_operator}, we can expand $\hat{H}_{\text{eff}}$ as a power series in $\lambda$:
\begin{equation}
    \hat{H}_{\text{eff}} = \sum_{M>0} \lambda^M \hat{H}^{(M)}_{\text{eff}}
\end{equation}
The $M$-th order term $\hat{H}^{(N)}_{\text{eff}}$ has the form
\begin{equation}
    \hat{H}^{(M)}_{\text{eff}}
    = \sum_{k_1 + \dots + k_{M-1} = M - 1, \, k_i \geq 0}
        f(k_i) \Proj_0 \hat{V} \Sproj^{k_1} \hat{V} \Sproj^{k_2} \hat{V} \cdots \hat{V} \Sproj^{k_{M-1}} \hat{V} \Proj_0,
    \label{eq:effective_hamiltonian_n_term}
\end{equation}
with appropriate coefficients $f(k_i)$.
The expression \eqref{eq:effective_hamiltonian_n_term} allows us to write down the effective Hamiltonian up to the desired order.

\subsection{Computation of the effective plaquette Hamiltonian}
\label{sub:computation_of_the_effective_plaquette_hamiltonian}

It is time to apply this approach to the plaquette circuit, whose Hamiltonian is
\begin{equation}
    \hat{H}_P
    = \hat{H}_0 + g\hat{V}_g + \lambda \hat{V}_{\lambda}
    = m\sum_{i} \hat{n}_i^2 + g \sum_{\link} \hat{n}_{ij}^2 - \lambda \sum_{\link} \cos(\hat{\phi}_i + \hat{\theta}_{ij} - \hat{\phi}_j).
    \label{eq:H_P_appendix}
\end{equation}
Given that we are interested in the regime where $m \gg g, \lambda$, we have set $\hat{H}_0 = m\sum_{i} \hat{n}_i^2 $ as the zeroth-order Hamiltonian and the rest as perturbation.
The space $\mathcal{H}_0$ is spanned by all the states where $n_i = 0$ for all $i = 1, \dots, 4$, which corresponds to the space of the ground states of $\hat{H}_0$.
Their energies are simply $E_0 = 0$.

Notice that the perturbation $V$ is made of two parts:
\begin{equation}
    \hat{V}_g = \sum_{\link} \hat{n}_{ij}^2
    \quad \text{and} \quad
    \hat{V}_{\lambda} = \sum_{\link} \cos(\hat{\phi}_i + \hat{\theta}_{ij} - \hat{\phi}_j).
\end{equation}
The former does not introduce any coupling between the space $\mathcal{H}_0$ and its complement inside the full Hilbert space, which means that $\Sproj \hat{V}_g \Proj_0 = 0$.
Therefore, from \eqref{eq:effective_hamiltonian_n_term}, it is clear that $\hat{V}_g$ only adds a correction at first order:
\begin{equation}
    H^{(1)}_{\text{eff}, g} = \Proj_0 \hat{V}_g \Proj_0 = \Proj_0 \qty( \sum_{\link} \hat{n}_{ij}^2  ) \Proj_0.
\end{equation}

The latter term $\hat{V}_{\lambda}$ is more interesting to analyze.
Consider a branch of the plaquette with two matter nodes and one gauge node.
In the charge basis, its states can be denoted with $\ket{n_i, n_{ij}, n_j}$.
The action of $\cosine(\hat{\phi}_i + \hat{\theta}_{ij} - \hat{\phi}_{j})$ maps $\ket{n_i, n_{ij}, n_j}$ into the superposition $ \frac{1}{2}(\ket{n_i + 1, n_{ij} + 1, n_j -1} + \ket{n_i - 1, n_{ij} - 1, n_j + 1})$.
Therefore, a state $\ket{\phi} \in \mathcal{H}_0$ (where $n_i = 0$ for all $i$) gets mapped into a virtual state $\ket{\zeta} \notin \mathcal{H}_0$ ($n_i \neq 0$ for some $i$).
Hence we can affirm that $\Proj_0 \hat{V} \Proj_0 = 0$.
From this, we can deduce that the only non-vanishing terms in \eqref{eq:effective_hamiltonian_n_term} appear for even order $M$.
In order to simplify notation, we define $\hat{U}_{ij} = e^{i(\hat{\phi}_i + \hat{\theta}_{ij} - \hat{\phi}_j)}$ such that $\hat{V}_{\lambda} = \frac{1}{2}\sum_{\link}(\hat{U}_{ij} + \hat{U}_{ij}^{\dagger})$.
At second order, we obtain
\begin{equation}
    H^{(2)}_{\text{eff}, \lambda}
    = \Proj_0 \hat{V}_{\lambda} \Sproj \hat{V}_{\lambda} \Proj_0
    = \Proj_0 \sum_{\link} \qty(
        \frac{1}{4} \hat{U}_{ij} \Sproj \hat{U}_{ij}^{\dagger} +
        \frac{1}{4} \hat{U}_{ij}^{\dagger} \Sproj \hat{U}_{ij}
      ) \Proj_0
\end{equation}
because the only non-vanishing terms are the ones where the cosine term is applied twice on the same link.
The expression above only produces a correction of $- \lambda^2 / (4m)$ for each branch to the ground state energy, resulting in a total correction of $\Delta E^{(2)} = - \lambda^2 / m$.

It becomes clear that to see any non-trivial dynamics, we need to move to the fourth order.
Given that $\Proj_0 \hat{V} \Proj_0 = 0$ and $S^0 = - \Proj_0$, we see that \eqref{eq:effective_hamiltonian_n_term} for $M = 4$ simplifies to \cite{takahashi1977half-filled}
\begin{equation}
    H^{(4)}_{\text{eff}, \lambda}
    = \Proj_0 \hat{V}_{\lambda} \Sproj \hat{V}_{\lambda} \Sproj \hat{V}_{\lambda} \Sproj \hat{V}_{\lambda} \Proj_0
    - \frac{1}{2} \qty(
        \Proj_0 \hat{V}_{\lambda} \Sproj^2 \hat{V}_{\lambda} \Proj_0 \hat{V}_{\lambda} \Sproj   \hat{V}_{\lambda} \Proj_0 +
        \Proj_0 \hat{V}_{\lambda} \Sproj   \hat{V}_{\lambda} \Proj_0 \hat{V}_{\lambda} \Sproj^2 \hat{V}_{\lambda} \Proj_0
    ).
    \label{eq:eff_H_lambda_4_order}
\end{equation}
In the same way as the previous case, the terms in brackets in \eqref{eq:eff_H_lambda_4_order} contribute only with a correction to the ground state energy.
With similar calculations, one finds that the resulting correction is $- \lambda^4 / (2 m^3)$.

\begin{figure}[b]
    \centering
    \includegraphics[]{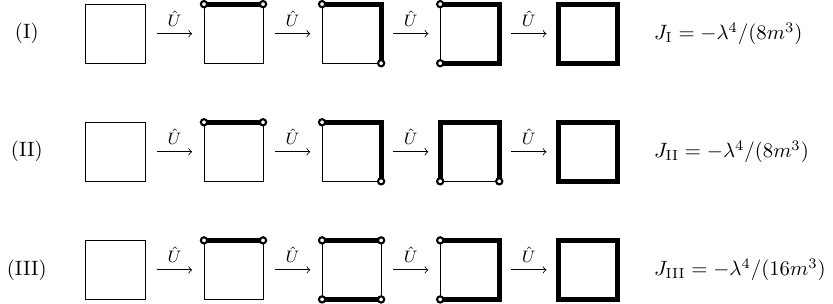}
    \caption{The three types of processes that contribute to the effective Hamiltonian $H^{(4)}_{\text{eff}, \lambda}$ and their effective energies $J$.
    The thick black lines stand for non-zero gauge charges, while the white dots represent non-zero matter charges.}
    \label{fig:perturbative_processes}
\end{figure}

On the other hand, the first term in \eqref{eq:eff_H_lambda_4_order} actually captures some non-trivial dynamics.
It shows that it is possible to connect a state $\ket{\varphi} \in \mathcal{H}_0$ to another state $\ket{\varphi^{\prime}} \in \mathcal{H}_0$ through three virtual states $\ket{\zeta_j} \notin \mathcal{H}_0$ ($j = 1, \dots, 3$).
In the expansion of the first term of \eqref{eq:eff_H_lambda_4_order}, only the terms with an even number of $\hat{U}_{ij}$ survive when sandwiched between the projectors $\Proj_0$.
Moreover, the factors that contain two $\hat{U}_{ij}$s and two $\hat{U}_{ij}^{\dagger}$ add at most a correction to the ground state energy.
We ignore them for the moment and focus only on the terms that introduce an effective plaquette interaction.
Given the consideration above, we write the expansion as follows:
\begin{equation}
    \Proj_0 \hat{V}_{\lambda} \Sproj \hat{V}_{\lambda} \Sproj \hat{V}_{\lambda} \Sproj \hat{V}_{\lambda} \Proj_0
    \simeq \frac{1}{16} \Proj_0 \sum_{\{\ev*{i j}\}} \left(
        \hat{U}_{i_1 j_1} \Sproj \hat{U}_{i_2 j_2} \Sproj \hat{U}_{i_3 j_3} \Sproj \hat{U}_{i_4 j_4} +
        \hat{U}_{i_1 j_1}^{\dagger} \Sproj \hat{U}_{i_2 j_2}^{\dagger} \Sproj \hat{U}_{i_3 j_3}^{\dagger} \Sproj \hat{U}_{i_4 j_4}^{\dagger}
    \right) \Proj_0
\end{equation}
where the sum is over all the possible combinations of links $ij$.
The only non-vanishing terms in the sum are the ones where each of the $\hat{U}$s (or $\hat{U}^{\dagger}$s) acts on a different site.
As a result, they give rise to an effective plaquette term $\hat{\square} = \cos(\hat{\theta}_{12} + \hat{\theta}_{24} - \hat{\theta}_{34} - \hat{\theta}_{13})$.
There are $4 \times 3 \times 2 = 24$ possible ways to order these $\hat{U}$, which we can classify into three types of processes, shown in Fig.~\ref{fig:perturbative_processes}.

Each type of process can start from any of the links and go in either clockwise or counterclockwise direction, meaning that for each type, we have 8 possible combinations.
The same reasoning applies to the second term, which involves $U_{ij}^{\dagger}$, as it is just the Hermitian conjugate of the first term.
Moreover, each process of type (I) and (II) have an effective energy of $J_{\text{I}, \text{II}} = - \lambda^4 / (8 m^3)$, while for the third type we have $J_{\text{III}} = - \lambda^4 / (16 m^3)$, because it goes through a virtual state with an extra pair of mass charges.
Combining all these data, we obtain an effective coupling
\begin{equation}
    J_{\square}
    = 2 \times \frac{1}{16} \qty(8 J_{\text{I}} + 8 J_{\text{II}} + 8 J_{\text{III}})
    = - \frac{\lambda^4}{m^3} \qty( \frac{1}{8} + \frac{1}{8} + \frac{1}{16}  )
    = - \frac{5 \lambda^4}{16 m^3}.
\end{equation}
Therefore, the resulting effective term at fourth order is
\begin{equation}
    \lambda^4 H^{(4)}_{\text{eff}, \lambda}
    = - \frac{5 \lambda^4}{16 m^3} \Proj_0
    \cos( \hat{\theta}_{12} + \hat{\theta}_{24} - \hat{\theta}_{34} - \hat{\theta}_{13} )
    \Proj_0
    + \Delta E^{(4)},
\end{equation}
where $\Delta E^{(4)}$ indicates correction to the ground state energy of the order $O(\lambda^4 / m^3)$.

In summary, the resulting effective Hamiltonian at fourth order is
\begin{equation}
    \hat{H}_{\text{eff}} =
    \Proj_0
    \qty(
        g \sum_{\link} \hat{n}_{ij}^2
        - \frac{5 \lambda^4}{16m^3}
        \cos( \hat{\theta}_{12} + \hat{\theta}_{24} - \hat{\theta}_{34} - \hat{\theta}_{13} )
    )
    \Proj_0
    + \Delta E^{(2)} + \Delta E^{(4)} + O(\lambda^5/m^4)
    \label{eq:Heff_4th_order}
\end{equation}

\subsection{Convergence of the perturbative series}
\label{sub:convergence_of_the_expansion}

The domain of validity of the effective Hamiltonian is dictated by the convergence radius of the expansion in \eqref{eq:resolvent_expansion}.
Thanks to Kato \cite{kato1949perturbation}, we can have a crude estimate of when the general series \eqref{eq:resolvent_expansion} is absolutely convergent.
In particular, the author finds
\begin{equation}
    \abs{\lambda} < \frac{d}{2 \norm*{\hat{V}}}
    \quad \text{or} \quad
    \norm*{\lambda \hat{V}} < \frac{d}{2},
\end{equation}
where, in our case, $d$ is the energy gap between the ground state and first excited state.

Because we are interested in the emerging plaquette interaction, we focus only on $\hat{V}_{\lambda}$.
The zeroth-order Hamiltonian $\hat{H}_0$ in \eqref{eq:H_P_appendix} has a gap $d = m$.
Furthermore, because the cosine is function bounded between $1$ and $-1$, we have that the norm of $\hat{V}$ is $\norm*{\hat{V}_{\lambda}} = 4$.
Therefore, the expansion is convergent when $\lambda < m/8$, which gives an estimate of the domain validity of the effective Hamiltonian \eqref{eq:Heff}.
In Appendix~\ref{app_sec:Numerical_results}, specifically in Fig.~\ref{fig:H_P_vs_H_eff_def}, we can see that at least for the lower spectrum, this condition is satisfied.

\section{Numerical results}
\label{app_sec:Numerical_results}

\subsection{Results for one plaquette circuit}
\label{app_subsec:Results}

\subsubsection{Choice of basis truncation for numerical analysis}
\label{app_subsubsec:Charge_vs_Mathie_basis}

\begin{figure}[t]
    \centering
    \includegraphics[scale = 0.25]{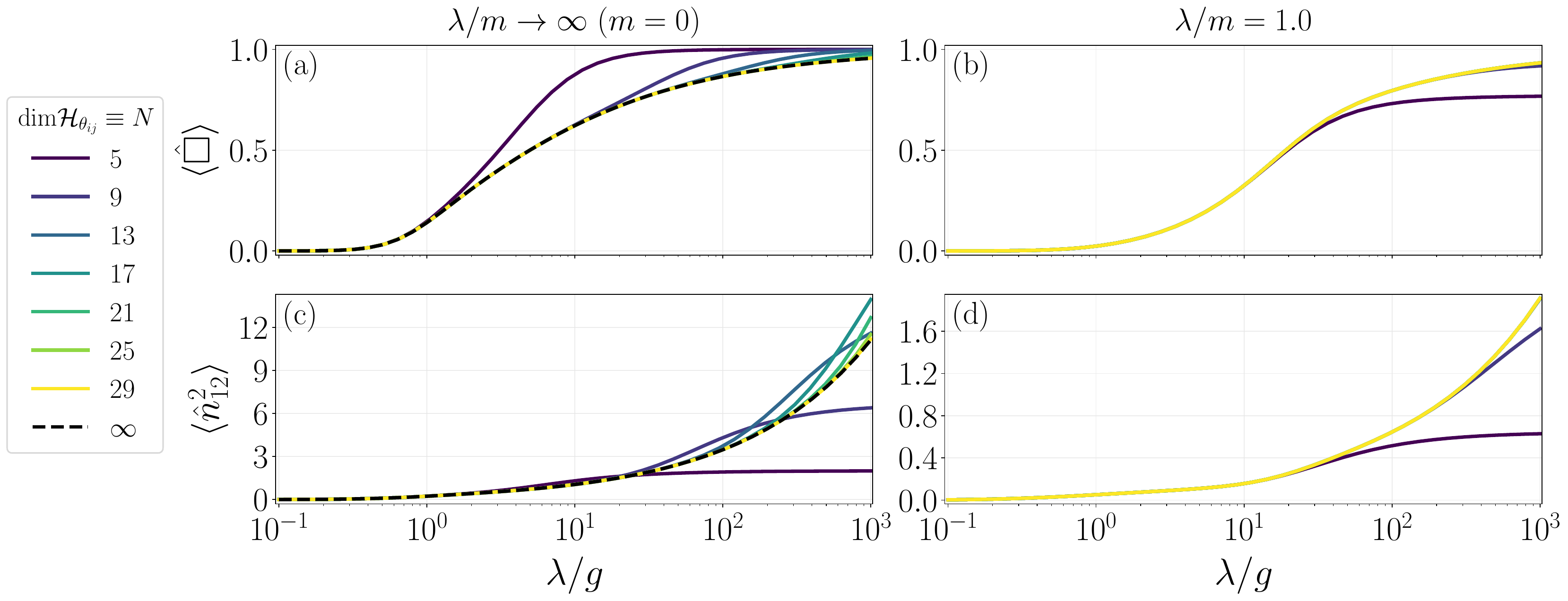}
    \caption{Ground-state properties of the plaquette Hamiltonian in the charge basis.
    (a---b) Expectation value of the plaquette operator $\hat{\square}$.
    (c---d) Expectation value of the $\hat{n}^2$ operator on a representative link.
    Left column: $\lambda/m \to \infty$ ($m=0$), where the system reduces to four independent charge qubits.
    Different truncations of the gauge charge basis are shown, and the analytical solution is indicated by the black dashed curve.
    A large local Hilbert space of dimension $N \gtrsim 25$ is required for accurate results over the full $\lambda/g$ range.
    Right column: $m\neq 0$ ($\lambda/m=1.0$), where the charge basis is more efficient.
    A smaller truncation, $N \gtrsim 9$, suffices.}
    \label{fig:validity_charge_basis}
\end{figure}

The plaquette Hamiltonian is most naturally expressed in the link (gauge charge) basis ${ \ket{n_{ij}} }$ for all $\langle ij \rangle \in \{12,24,13,34\}$. 
This discrete basis has integer eigenvalues (in units of $2e$), representing the number of charges at each circuit node, following $\hat{n}_{ij} \ket{n_{ij}} = n_{ij} \ket{n_{ij}}$. 
The operators $e^{\pm i \hat{\theta}_{ij}}$ act as raising and lowering operators, $e^{\pm i \hat{\theta}_{ij}} \ket{n_{ij}} = \ket{n_{ij} \pm 1}$. 
In matrix form, this reads $\hat{n}_{ij} = \sum_{n_{ij} \in \mathbb{Z}} n_{ij} \, \lvert n_{ij} \rangle \langle n_{ij} \rvert$ and $e^{\pm i\hat{\theta}_{ij}} = \sum_{n_{ij} \in \mathbb{Z}} \lvert n_{ij} \pm 1 \rangle \langle n_{ij} \rvert$.  Recalling \eqref{eq:gauge_kinetic_term_matrix_form}, the plaquette Hamiltonian can be split as $\hat{H}_{\text{P}}(m,g,\lambda) = \hat{K}(m, g) + \hat{V}(\lambda)$, with
\begin{equation}
    \begin{split}
        \hat{K}(m, g)    & \equiv m\sum_{i} \hat{n}_i^2 + g \sum_{\link} \hat{n}_{ij}^2 =
    \begin{pmatrix}
        \hat{n}_{12} & \hat{n}_{24} & \hat{n}_{13} & \hat{n}_{34}
    \end{pmatrix}
    \begin{pmatrix}
        g + 2m & -m      & m   &    \\
        -m      & g + 2m &      & m \\
        m     &        & g+2m & -m  \\
        & m     & -m    & g+2m
    \end{pmatrix}
    \begin{pmatrix}
        \hat{n}_{12} \\ \hat{n}_{24} \\ \hat{n}_{13} \\ \hat{n}_{34}
    \end{pmatrix} \\
        \hat{V}(\lambda) & \equiv - \lambda \sum_{\link} \cos(\hat{\phi}_i + \hat{\theta}_{ij} - \hat{\phi}_j) = - \lambda \sum_{\link} \cos \hat{\theta}_{ij}  = -\frac{\lambda}{2} \sum_{\link} \left( e^{i \hat{\theta}_{ij}} + e^{-i \hat{\theta}_{ij}}  \right).
        \label{eq:plaq_hamiltonian_divided_3}
    \end{split}
\end{equation}
To perform a reliable numerical study, the appropriate truncation of the local Hilbert space must be determined. 
The charge basis becomes suboptimal at large $\lambda/m$ or $\lambda/g$, where cosine terms dominate. In the limit $\lambda/m \to \infty$, the system has a known analytical solution (Appendix~\ref{app_sec:Mathieu}). 
We therefore benchmark the truncation by computing the ground-state expectation value of the plaquette operator and the corresponding electric-field fluctuations as a function of $\lambda/g$ [see Fig.~\ref{fig:validity_charge_basis}]. Capturing the static properties at large $\lambda/g$ requires a large local Hilbert space of dimension $N \gtrsim 25$. 
For moderate values of $\lambda/m$ and $\lambda/g$, a smaller truncation suffices. 
Similarly, reducing $\lambda/m$ also allows for a smaller local basis without compromising accuracy.

\begin{figure}[t]
    \centering
    \includegraphics[scale = 0.25]{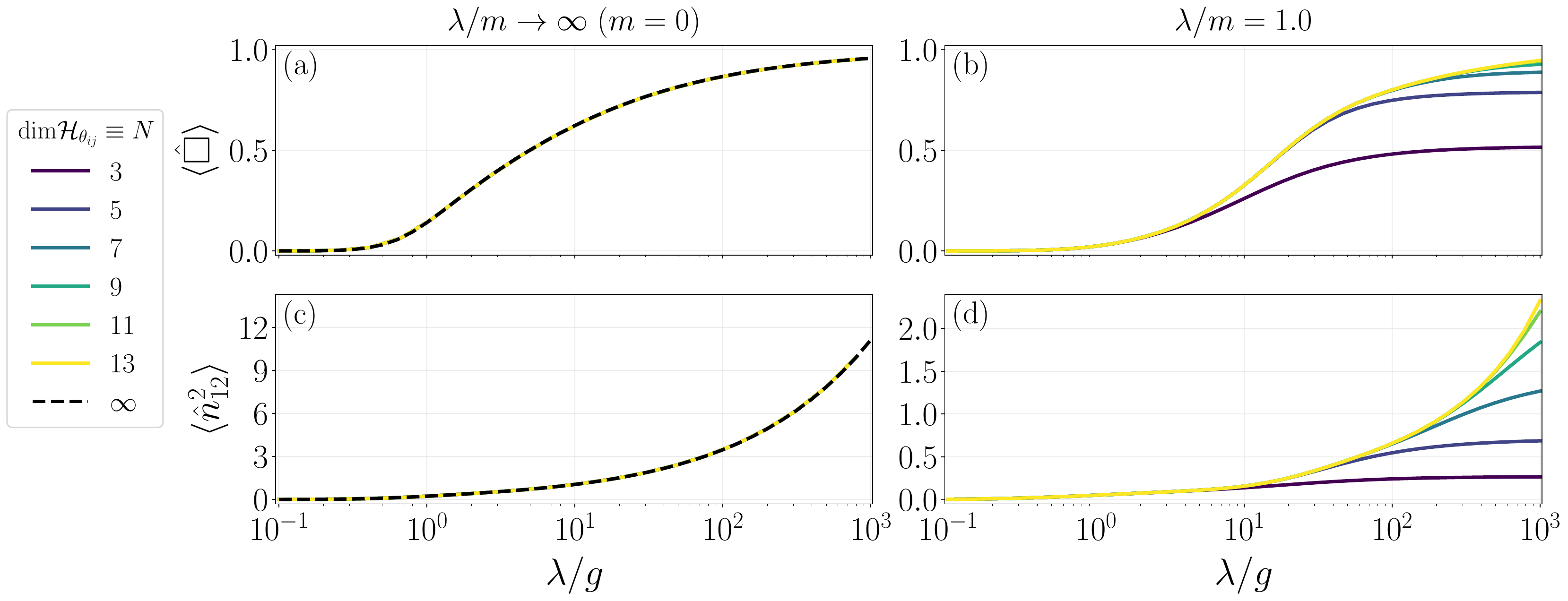}
    \caption{Ground-state properties of the plaquette Hamiltonian using the Mathieu basis.
    (a--b) Expectation value of the plaquette operator $\hat{\square} = \cos{(\hat{\theta}_{12} + \hat{\theta}_{24} - \hat{\theta}_{13} - \hat{\theta}_{34})}$.
    (c--d) Expectation value of the $\hat{n}^2$ operator on a representative link.
    Left column: $\lambda/m \to \infty$ ($m=0$), where the Hamiltonian is diagonal, and the Mathieu basis trivially reproduces the analytical solution.
    Right column: $m\neq 0$ ($\lambda/m = 1.0$), where the off-diagonal couplings appear and a larger local Hilbert space of dimension $N \gtrsim 13$, is needed for convergence.
    }
    \label{fig:validity_mathieu_basis}
\end{figure}

We now consider an alternative basis. Rewriting the plaquette Hamiltonian as $\hat{H}_{\rm P} = \hat{H}_{\rm links}(m, g, \lambda) + H_{\rm mass}(m) $ as in Appendix~\ref{app_sec:Mathieu}, leaves
\begin{equation}
    \begin{split}
        \hat{H}_{\rm links}(m, g, \lambda)    & \equiv \sum_{\langle ij \rangle} \hat{H}_{ij }= \sum_{\langle ij \rangle}(g + 2m)\hat{n}_{ij}^2 -\lambda \cos\hat{\theta}_{ij}, \  \langle ij \rangle \in \{ 12,24,13,34 \}\\
        \hat{H}_{\rm mass}(m) & \equiv 2m \left(\hat{n}_{12} \hat{n}_{13} + \hat{n}_{24} \hat{n}_{34} - \hat{n}_{24} \hat{n}_{12} - \hat{n}_{34} \hat{n}_{13} \right).
        \label{eq:plaq_hamiltonian_divided_2}
    \end{split}
\end{equation}
Each link Hamiltonian $\hat{H}_{ij}$ is exactly solvable, with eigenstates and eigenvalues determined by Mathieu’s differential equation for $q = -2\lambda/(2m+g)$,
\begin{equation}
    \begin{split}
        &\ket{\mathcal{M}}\in \{ \ket{{\rm ce}_{0}}, \ket{{\rm se}_{2}}, \ket{{\rm ce}_{2}}, \ket{{\rm se}_{4}}, \ket{{\rm ce}_{4}},\dots \},\\
        &E_{\mathcal{M}} = \frac{2m+g}{4} a_{\mathcal{M}}(q),
    \end{split}
\end{equation}
where $a_{\mathcal{M}} \in \{a_{0}, b_2, a_2, b_4, a_4, \dots\}$. 
Truncating to $N$ Mathieu states per link yields a plaquette Hilbert space of dimension $N^4$. 
On this basis, $\hat{H}_{ij}$ is diagonal with analytically known elements, while $\hat{H}_{\rm mass}$ introduces non-diagonal couplings. 
To preserve hermiticity numerically, we symmetrize the charge operator, $\hat{n}_{ij} \to (\hat{n}_{ij} + \hat{n}_{ij}^{\dagger})/2$. 
Reproducing Fig.~\ref{fig:validity_charge_basis} in this basis recovers the exact analytical curve in the $\lambda/m \to \infty$ limit independently of $N$, as the Hamiltonian is diagonal [see Fig.~\ref{fig:validity_mathieu_basis}]. 
Using this Mathieu basis at high $\lambda/m$ with $m\neq0$ allows for a smaller local Hilbert space than the charge basis while maintaining accuracy.

\subsubsection{Validity of the effective Hamiltonian and plaquette interaction visualization}
\label{app_subsubsec:Validity_effective_Hamiltonian}

Recall the effective plaquette Hamiltonian \eqref{eq:Heff_4th_order} (without the energy corrections)
\begin{equation}
    \hat{H}_{\text{eff}} =
    \Proj_0
    \qty(
        g \sum_{\link} \hat{n}_{ij}^2
        - \frac{5 \lambda^4}{16m^3}
        \cos( \hat{\theta}_{12} + \hat{\theta}_{24} - \hat{\theta}_{34} - \hat{\theta}_{13} )
    )
    \Proj_0.
    \label{eq:Heff}
\end{equation}

\begin{figure}[b]
    \centering
    \includegraphics[scale = 0.53]{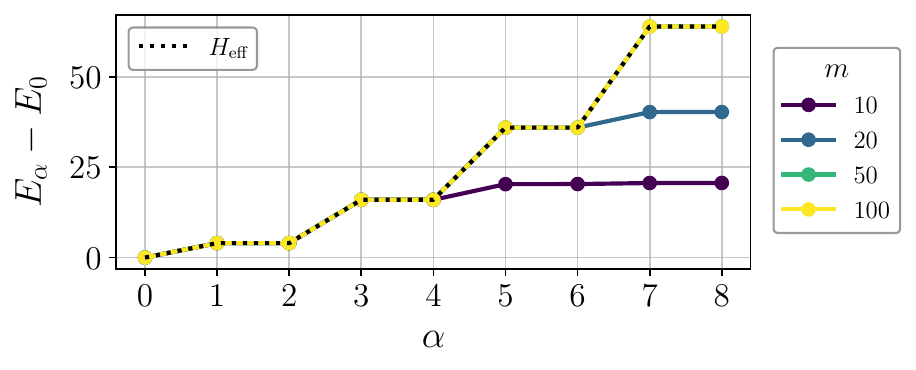}
    \caption{Numerical comparison between the effective and full Hamiltonian. Isolation of the $\lambda$ perturbation by fixing $\lambda = g = 1$ and varying $m$. Low-energy spectrum of the full Hamiltonian (first nine eigenvalues with index $\alpha$), shifted to zero energy. The effective Hamiltonian spectrum corresponds to the dotted black line. Increasing $m$ enhances the static-matter regime, strengthening the validity of the effective description in the low-energy sector and increasing the number of eigenstates that coincide with the full spectrum. Overall, in the proper static-matter regime, the effective Hamiltonian spectrum reproduces the full one with relative errors of order $\sim 10^{-7}$. Calculations are performed in the charge basis with local Hilbert space of dimension $N = 25$.}
    \label{fig:H_P_vs_H_eff_def}
\end{figure}

As discussed in Appendix~\ref{app_sec:Heff_static_m}, this perturbative description is valid in the regime $m \gg \lambda,g$, where the low-energy sector is characterized by $n_i = 0$ on all sites for sufficiently small gauge-charge occupation numbers $n_{ij}$. 
In this limit, the dynamics is governed by an effective plaquette interaction involving the four gauge flux variables $\hat{\theta}_{ij}$, generated through virtual transitions to high-energy states. 
Moreover, an estimate of the domain of validity, determined by focusing on the $\lambda$ perturbation, yielded $m > 8\lambda$. 
Fig. \ref{fig:H_P_vs_H_eff_def} compares the spectrum of the effective plaquette Hamiltonian with that of the full Hamiltonian \eqref{eq:H_P_appendix}. 
In the regime $m \gg \lambda \sim g $, the two agree remarkably well in the low-energy sector, with relative errors of order $\sim 10^{-7}$, consistent with the predicted domain of validity.

\begin{figure}[t]
    \centering
    \includegraphics[width=\textwidth]{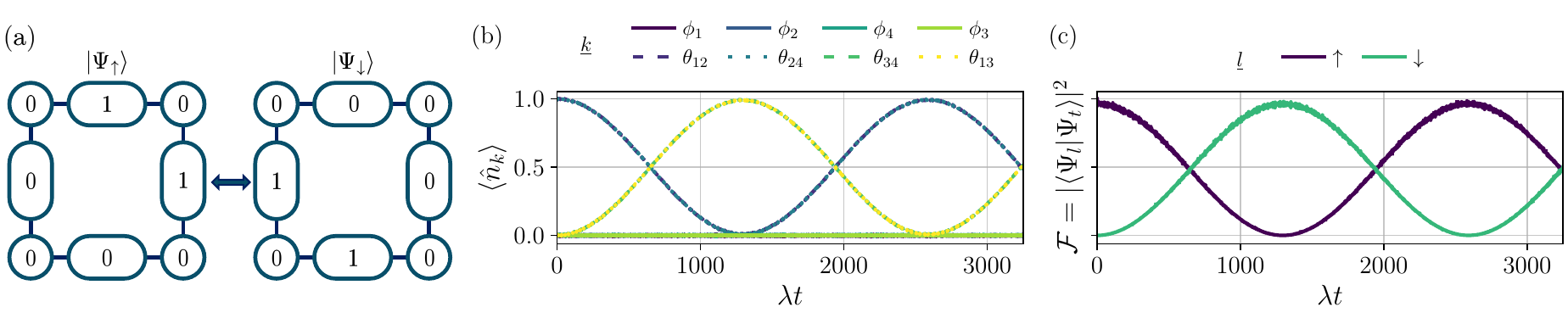}
    \caption{Dynamics in the static matter regime.  (a) ``UP" $\ket{\Psi_{\uparrow}}\equiv\ket{1, 1, 0, 0}$ and ``DOWN" $\ket{\Psi_{\downarrow}}=\ket{0, 0, 1, 1}$ configurations, expressed in the gauge charge basis $\{ \ket{ n_{12},  n_{24}, n_{13}, n_{34} } \}$. The Gauss's law static charges are chosen as $q_{1} = q_{4} = 1$ and $q_{2}= q_{3} = 0$, ensuring vanishing matter occupation. The two states are energetically degenerated and are connected by the action of the plaquette operator. (b) Time evolution of the expectation values of charge number operators, initialized in $\ket{\Psi_{t=0}}=\ket{\Psi_{\uparrow}}$, for parameters $m = 5$, $\lambda = 1$, $g = 0.2$ $(\lambda / m = 0.2, \lambda / g = 5)$, using a local Hilbert space of dimension $N = 11$. Only the gauge d.o.f exhibit coherent oscillations, while matter occupations remain frozen at zero, confirming the static matter regime $m \gg \lambda \gtrsim g $. (c) Fidelity of the time-evolved state in (b) with respect to $\ket{\Psi_{\uparrow}}$ and $\ket{\Psi_{\downarrow}}$ as a function of time, showing coherent oscillations consistent with the population dynamics. An equivalent behavior is obtained when initializing the system in $\ket{\Psi_{\downarrow}}$. }
    \label{fig:up_down_t_evol}
\end{figure}

The perturbative interaction can also be visualized dynamically through coherent population exchange $\Delta n_{ij} = \pm 1$ between the pairs ($\hat{n}_{12},\hat{n}_{24}$) and ($\hat{n}_{13},\hat{n}_{34}$). 
To this end, we define two plaquette configurations in the gauge-charge basis, denoted ``UP" and ``DOWN", $\ket{\Psi_{\uparrow}}\equiv\ket{1, 1, 0, 0}$ and $\ket{\Psi_{\downarrow}}=\ket{0, 0, 1, 1}$, respectively [see Fig.~\ref{fig:up_down_t_evol}(a)], reflecting the location of charge on the upper or lower links. Initializing the system in either state and evolving it in time reveals clear oscillations of the charge number populations, as shown in Fig.~\ref{fig:up_down_t_evol}(b--c). 
These oscillations persist in the presence of shunting capacitances in the Josephson junctions, since the additional terms introduced by circuit quantization (see Appendix~\ref{app_sec:Circuit_quantization}) preserve gauge invariance.

In the static matter regime, the low-energy dynamics are therefore governed by coherent oscillations, with an effective angular frequency $ \omega \equiv |J_{\square}| = 5 \lambda^4 / (16 m^3)$, as predicted by perturbation theory. 
The oscillation frequency is extracted numerically by Fourier transforming the time-dependent expectation value of $\hat{n}_{12}$ after subtracting its mean value. 
The dominant spectral peak yields the frequency, which can be visually cross-checked against the period of oscillations. 
While finite-size truncation effects induce small shifts in the numerical value, with relative errors of order $\sim 10^{-3}$, the agreement with the perturbative prediction remains robust.

Experimentally, $\lambda$ would initially be tuned to small values to suppress the vortex dynamics. An initial state can then be prepared in the gauge subspace using standard Rabi-like control pulses on the link islands. 
Nonadiabatically turning on $\lambda$ will result in coherent vortex dynamics. After a variable time, the system can again be returned nonadiabatically to the ``frozen" state, arresting the vortex dynamics, at which point the charge states can be read out.

\subsection{Extension to ($2 \times 2$)-plaquette lattice}
\label{app_subsec:2x2_plaquette}

Tensor network methods are well-suited for characterizing quantum states that satisfy the area law of entanglement, such as the low-energy eigenstates of Hamiltonians with local interactions~\cite{perez2006matrix,montangero2018introduction,RevModPhys.93.045003}. 
Recently, they have also been successfully applied to superconducting circuits in the context of quantum computation and quantum simulation~\cite{weiss2019spectrum,di2021efficient,roy2023quantum,rui2025tensor}. In this section, we present results on the ground-state properties of a $(2\times2)$-plaquette lattice obtained with the matrix product state implementation of the density matrix renormalization group (DMRG) method.

Let us first spell out the Hamiltonian for the system. In this case, there are $9$ matter nodes and $12$ gauge nodes, see Fig.~\ref{fig:sketch_2x2}. 
Ignoring the stray capacitances and eliminating the matter nodes as in Appendix~\ref{app:Gauss_match_dof}, we find the $(2\times2)$-plaquette Hamiltonian to be $\hat{H}_{\text{4P}}(m,g,\lambda) = \hat{K}(m, g) + \hat{V}(\lambda)$, where $\hat{K}(m, g) = g\hat{\boldsymbol{n}}^T\hat{\boldsymbol{n}}+m\hat{\boldsymbol{n}}^T_{\mathrm{v}}M_{\mathrm{v}}\hat{\boldsymbol{n}}_{\mathrm{v}}$, with the coefficient matrix
\begin{equation}
    M_{\mathrm{v}} =
    \begin{pmatrix}
        2 & -1 & 1 & -1 &  &  &  &  &  &  & &  \\
        -1 & 2 &  & 1 & -1 &  &  &  &  &  & &  \\
        1 &  & 2 &  &  & -1 &  & -1 &  &  & &  \\
        -1 & 1 &  & 2 &  & 1 & -1 &  & -1 & &  &  \\
         & -1 &  &  & 2 &  & 1 &  &  & -1 & &  \\
         &  & -1 & 1 &  & 2 & -1 & 1 & -1 &  & &  \\
         &  &  & -1 & 1 & -1 & 2 &  & 1 & -1 &  & \\
         &  & -1 &  &  & 1 &  & 2 &  &  & -1 &  \\
         &  &  & -1 &  & -1 & 1 &  & 2 &  & 1 & -1 \\
         &  &  &  & -1 &  & -1 &  &  & 2 &  & 1 \\
         &  &  &  &  &  &  & -1 & 1 &  & 2 & -1 \\
         &  &  &  &  &  &  &  & -1 & 1 & -1 & 2
    \end{pmatrix}
    \label{eq:gauge_kinetic_term_mass_coeff_4P}
\end{equation}
and the gauge charge operator vector
\begin{equation}
    \hat{\boldsymbol{n}}^T_{\mathrm{v}} = \begin{pmatrix}
        \hat{n}_{12},\, \hat{n}_{23},\, \hat{n}_{14},\, \hat{n}_{25},\, \hat{n}_{36},\, \hat{n}_{45},\, \hat{n}_{56},\, \hat{n}_{47},\, \hat{n}_{58},\, \hat{n}_{69},\, \hat{n}_{78},\, \hat{n}_{89}
    \end{pmatrix}.
    \label{eq:nv_vec_4P}
\end{equation}
Here, the diagonal elements are all $2$. If two gauge links do not share a matter vertex, the corresponding off-diagonal element vanishes; otherwise, the off-diagonal element is $1$ if the links has the same origin or destination (e.g., links $12$ and $14$, or links $25$ and $45$), and $-1$ if the shared vertex is the origin of one link and the destination of the other (e.g., links $12$ and $23$).
The potential term $\hat{V}(\lambda)$ takes the familiar form where all gauge links in \eqref{eq:nv_vec_4P} appear in the sum,
\begin{equation}
    \hat{V}(\lambda) = -\lambda \sum_{\link}  \cos(\hat{\theta}_{ij}).
\end{equation}

\begin{figure}[t]
    \centering
    \includegraphics[scale = 0.26]{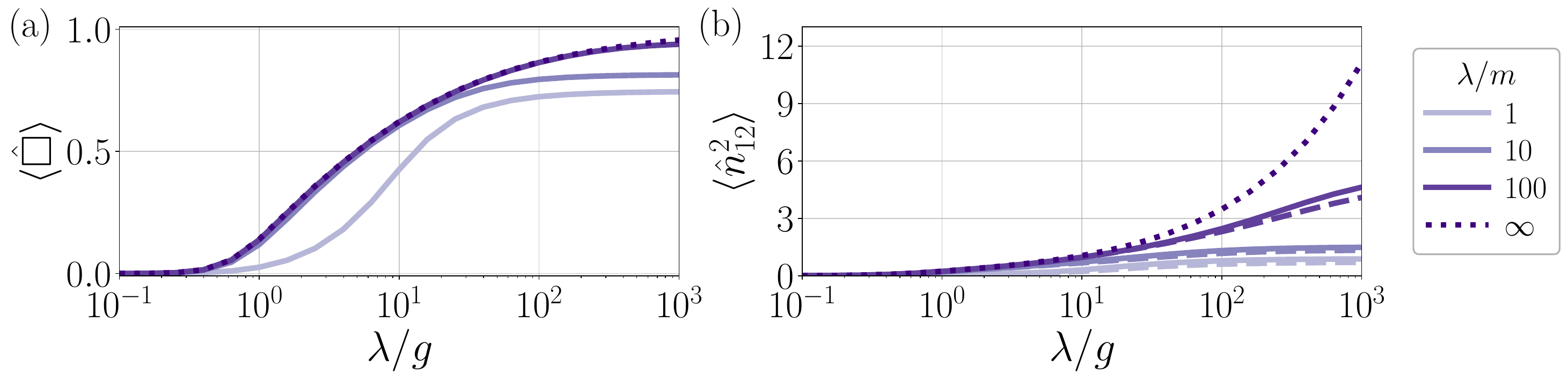}
    \caption{Ground-state properties of a $\mathrm{(2\times2)}$-plaquette lattice obtained using DMRG. (a) Expectation value of the plaquette operator $ \langle \hat{\square}\rangle$, identical for all four plaquettes in the ground state. (b) Electric-field fluctuations $\langle \hat{n}^2 \rangle$ on the four inner links (e.g. $\langle \hat{n}_{25}^2 \rangle$, solid) and the eight outer links (e.g. $\langle \hat{n}_{12}^2 \rangle$, dashed). Both are shown as functions of $\lambda / g$ and $\lambda / m$. The dotted curve represents in both plots the theoretical limit for $\lambda/m = \infty $ ($m=0$). 
    Mathieu basis restricted to the local Hilbert space of dimension $N = 5$.}
    \label{fig:TN_2x2}
\end{figure}

DMRG calculations were performed using the two-site algorithm in the TeNPy Library~\cite{tenpy2024}, and the results for the single-plaquette system are in excellent agreement with Fig.~\ref{fig:GS_observables}. 
In Fig.~\ref{fig:TN_2x2}, we present the results for the $(2\times2)$-plaquette lattice, keeping $N=5$ lowest-energy states in the Mathieu basis. 
We observe the same qualitative behavior as in Fig.~\ref{fig:GS_observables}: for fixed $\lambda$ and $g$, increasing $m$ suppresses both $ \langle \hat{\square}\rangle$ (identical for all four plaquettes) and $\langle \hat{n}^2 \rangle$ (here taking different values for links $12$ and $25$). 
However, as pointed out in Sec.~\ref{app_subsubsec:Charge_vs_Mathie_basis}, $N=5$ is likely insufficient for numerical convergence; this is especially clear in the $\lambda/g \to \infty$ case, where $ \langle \hat{\square}\rangle$ for $m\neq0$ does not approach the $m=0$ values as in Fig.~\ref{fig:GS_observables}. 
We leave the detailed numerical characterization of larger circuits for future work.

\begin{figure}[t]
    \centering
    \includegraphics[scale = 0.23]{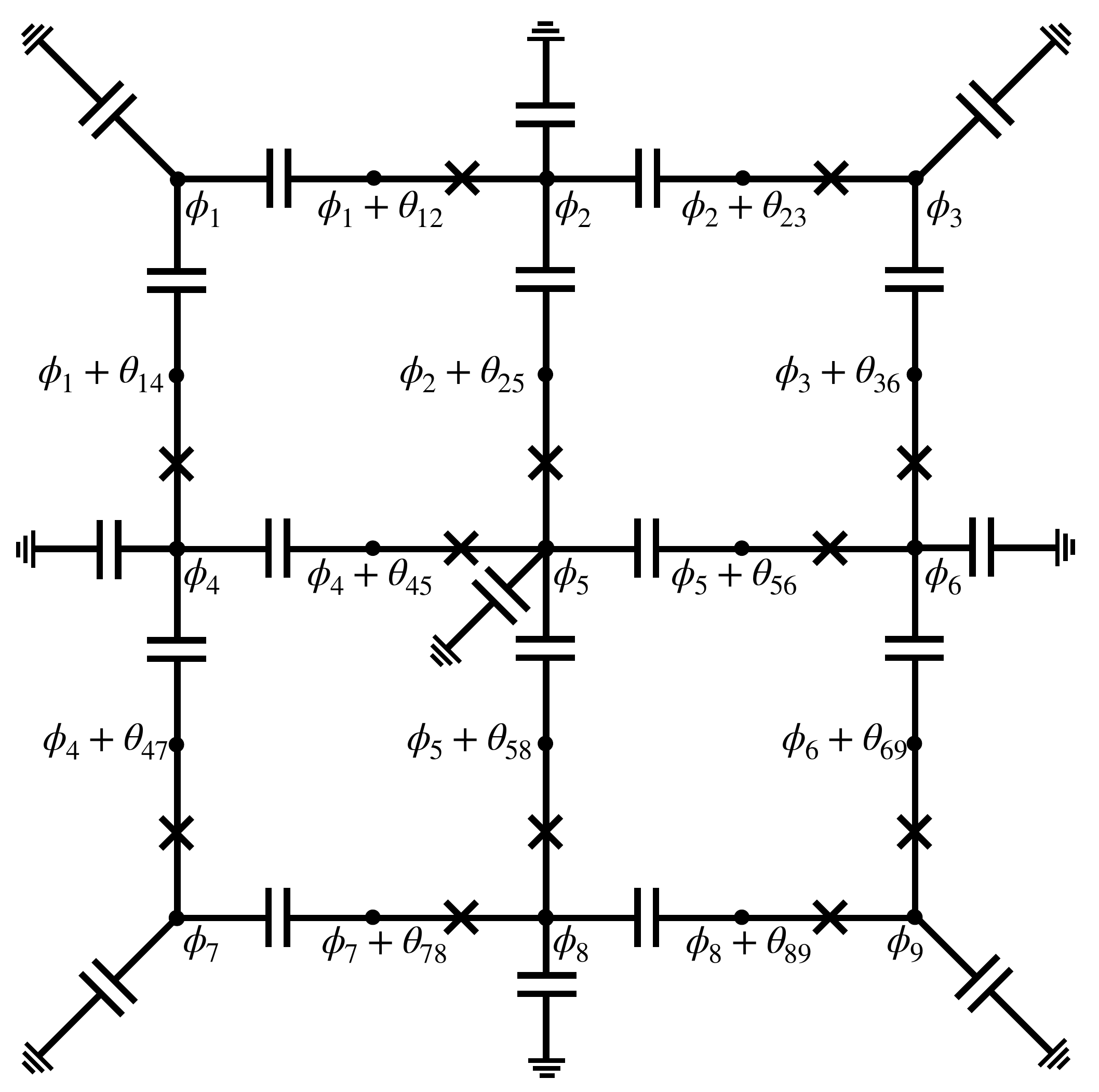}
    \caption{Sketch of a $(2\times2)$-plaquette lattice.}
    \label{fig:sketch_2x2}
\end{figure}

\end{document}